\documentclass{aastex631}

\newcommand{\hr}{\texttt{HR5}}
\newcommand{\pgalf}{\texttt{PGalF}}

\definecolor{red}{RGB}{250,0,0}

\shortauthors{Yoo et al.}

\graphicspath{{./}{figures/}}

\begin{document}

\title{Spatial Distribution of Intracluster Light versus Dark Matter in Horizon Run 5}
\shorttitle{Spatial Distribution of ICL vs DM}

\correspondingauthor{Cristiano G. Sabiu}
\email{csabiu@uos.ac.kr}

\author[0000-0002-6841-8329]{Jaewon Yoo}
\affiliation{Quantum Universe Center, Korea Institute for Advanced Study, 85 Hoegi-ro, Dongdaemun-gu, Seoul 02455, Korea}

\author[0000-0001-9521-6397]{Changbom Park}
\affiliation{School of Physics, Korea Institute for Advanced Study, 85 Hoegiro, Dongdaemun-gu, Seoul 02455, Korea}

\author[0000-0002-5513-5303]{Cristiano G. Sabiu}
\affiliation{Natural Science Research Institute (NSRI), University of Seoul, Seoul 02504, Korea}

\author[0000-0001-5427-4515]{Ankit Singh}
\affiliation{School of Physics, Korea Institute for Advanced Study, 85 Hoegiro, Dongdaemun-gu, Seoul 02455, Korea}

\author[0000-0002-9434-5936]{Jongwan Ko}
\affiliation{Korea Astronomy and Space Science Institute (KASI), Daedeokdae-ro, Daejeon 34055, Korea}
\affiliation{University of Science and Technology (UST), Gajeong-ro, Daejeon 34113, Korea}

\author[0000-0002-6810-1778]{Jaehyun Lee}
\affiliation{Korea Astronomy and Space Science Institute (KASI), Daedeokdae-ro, Daejeon 34055, Korea}
\affiliation{School of Physics, Korea Institute for Advanced Study, 85 Hoegiro, Dongdaemun-gu, Seoul 02455, Korea}

\author[0000-0003-0695-6735]{Christophe Pichon}
\affiliation{School of Physics, Korea Institute for Advanced Study, 85 Hoegiro, Dongdaemun-gu, Seoul 02455, Korea}
\affiliation{CNRS and Sorbonne Université, UMR 7095, Institut d’Astrophysique de Paris, 98 bis, Boulevard Arago, F-75014 Paris, France}
\affiliation{IPhT, DRF-INP, UMR 3680, CEA, L’Orme des Merisiers, Bât 774, F-91191 Gif-sur-Yvette, France}

\author[0000-0002-5751-3697]{M. James Jee}
\affiliation{Department of Astronomy, Yonsei University, 50 Yonsei-ro, Seoul 03722, Korea}
\affiliation{Department of Physics, University of California, Davis, One Shields Avenue, Davis, CA 95616, USA}

\author[0000-0003-4446-3130]{Brad K. Gibson}
\affiliation{Astrophysics Group, Keele University, Keele, Staffordshire, ST5 5BG, UK}

\author{Owain Snaith}
\affiliation{University of Exeter, School of Physics and Astronomy, Stocker Road, Exeter, EX4 4QL, UK}

\author[0000-0002-4391-2275]{Juhan Kim}
\affiliation{Center for Advanced Computation, Korea Institute for Advanced Study, 85 Hoegiro, Dongdaemun-gu, Seoul 02455, Korea}

\author[0000-0001-5135-1693]{Jihye Shin}
\affiliation{Korea Astronomy and Space Science Institute (KASI), Daedeokdae-ro, Daejeon 34055, Korea}

\author[0000-0003-4164-5414]{Yonghwi Kim}
\affiliation{Korea Institute of Science and Technology Information, 245 Daehak-ro, Yuseong-gu, Daejeon, 34141, Korea}

\author[0000-0003-4032-8572]{Hyowon Kim}
\affiliation{Korea Astronomy and Space Science Institute (KASI), Daedeokdae-ro, Daejeon 34055, Korea}
\affiliation{University of Science and Technology (UST), Gajeong-ro, Daejeon 34113, Korea}

\begin{abstract}
One intriguing approach for studying the dynamical evolution of galaxy clusters is to compare the spatial distributions among various components, such as dark matter, member galaxies, gas, and intracluster light (ICL). 
Utilizing the recently introduced Weighted Overlap Coefficient (WOC) \citep{2022ApJS..261...28Y}, we analyze the spatial distributions of components within 174 galaxy clusters ($M_{\rm tot}> 5 \times 10^{13} M_{\odot}$, $z=0.625$) at varying dynamical states in the cosmological hydrodynamical simulation Horizon Run 5. We observe that the distributions of gas and the combination of ICL with the brightest cluster galaxy (BCG) closely resembles the dark matter distribution, particularly in more relaxed clusters, characterized by the half-mass epoch. 
The similarity in spatial distribution between dark matter and BCG+ICL  mimics the changes in the dynamical state of clusters during a major merger.  Notably, at redshifts $>$ 1, BCG+ICL traced dark matter more accurately than the gas.
Additionally, we examined the one-dimensional radial profiles of each component, which show that the BCG+ICL is a sensitive component revealing the dynamical  state of clusters. 
We propose a  new method that can approximately recover the dark matter profile by scaling the BCG+ICL radial profile.
Furthermore, we find a recipe for tracing dark matter in unrelaxed clusters by including the most massive satellite galaxies together with BCG+ICL distribution. Combining the BCG+ICL and the gas distribution enhances the dark matter tracing ability. 
Our results imply that the BCG+ICL distribution is an effective tracer for the dark matter distribution, and the similarity of spatial distribution may be a useful probe of the dynamical state of a cluster.

\end{abstract}


\section{Introduction} \label{sec:intro}
Galaxy clusters lie at a unique crossroads in the field of astrophysics and cosmology, enabling us to probe the large-scale structure formation and the influences of dark matter and dark energy on the expansion history of the universe \citep{voit2005, kravtsov2012}. 
Examining cluster properties that reflect their evolutionary state offers valuable insight into the formation process of clusters. Among these properties, the phenomenon of \textit{intracluster light} (ICL), which refers to the luminosity emitted by stars not tethered to any specific cluster member galaxy, has emerged as a promising\
means to unravel the assembly history of galaxy clusters \citep[][and \cite{2021Galax...9...60C} for a recent review]{1951PASP...63...61Z, 1998Natur.396..549G, 2002ApJ...575..779F, 2004ApJ...617..879L, 2005ApJ...618..195G, 2005MNRAS.358..949Z, 2005ApJ...631L..41M,  2017ApJ...834...16M, 2018MNRAS.474.3009D, 2018ApJ...862...95K, 2019A&A...622A.183J, 2021ApJ...910...45M, 2021MNRAS.508.2634Y}. The ICL comprises a considerable fraction of stars, ranging from 0 to 40 percent of the total cluster light \citep{2015MNRAS.449.2353B, 2018ApJ...857...79J, 2018MNRAS.474..917M, 2021MNRAS.508.2634Y, 2022NatAs...6..308M, 2023Natur.613...37J}. Studies using simulations have revealed that various dynamical interactions among galaxies within the cluster environment, such as violent relaxation after major merging, tidal stripping, and tidal disruption of galaxies, contribute to the ICL \citep{2007MNRAS.377....2M, 2007ApJ...666...20P, 2007ApJ...668..826C, 2010MNRAS.406..936P, 2011ApJ...732...48R,  2015MNRAS.451.2703C,2014MNRAS.437.3787C,2018MNRAS.479..932C}.

A critical aspect of the study of galaxy cluster evolution is understanding how the spatial distribution of dark matter varies with factors such as redshift, mass, and the dynamical state of the cluster. Gravitational weak-lensing is an effective method of studying the dark matter distribution within clusters \citep{1993ApJ...404..441K,2006ApJ...648L.109C,2007ApJ...661..728J,2014ApJ...784...90O}. However, such studies require the identification of a large number of background galaxies and accurate measurement of their shape, which is quite challenging.
Consequently, additional quantities that can reveal the dark matter distribution and provide supplementary information on clusters are desirable. 

Like dark matter, ICL is collisionless and gravitationally bound to the overall gravitational potential of its host cluster rather than to that of individual galaxies. Within the $\Lambda$CDM cosmological framework, 
galaxy clusters grow through the hierarchical merging of smaller structures, and it is reasonable to expect that  such merging events can enrich the ICL.
Prior works have demonstrated that the ICL traces the dark matter distribution well \citep{2019MNRAS.482.2838M, 2020MNRAS.494.1859A, 2022ApJS..261...28Y}, 
and the properties of ICL are closely related to the evolutionary state of galaxy clusters. Therefore,  comparing the spatial distributions of ICL and dark matter could   prove to be a critical step toward utilizing ICL as a \textit{luminous dark matter tracer} and a \textit{probe of the evolutionary state of galaxy clusters}.

We apply a novel approach to assess the similarity between two-dimensional spatial distributions, denoted as the Weighted Overlap Coefficient \citep[WOC;][]{2022ApJS..261...28Y} method. 
Our study compares two spatial distributions without assuming any specific relationship between their respective signal strengths.
The WOC method computes the overlap fraction between two distributions across various density threshold levels while incorporating signal strength weighting and normalizing similarity measures to a range of 0 to 1. The method is non-parametric, intuitive, and avoids numerical errors associated with fitting, enabling robust similarity quantification under different smoothing factors, centers, and binning choices. Additionally, WOC performs well for disconnected contours and masked maps with multiple significant sub-structures, as exemplified by the Bullet cluster and Coma cluster undergoing active merging events. Unlike other methods, WOC does not require computing individual contours, reducing bias in masked maps.

To explore the ICL, which comprises unbound stars within galaxy clusters, leveraging high-fidelity numerical simulations is crucial. The Horizon Run 5 (\hr) stands as a significant cosmological hydrodynamical simulation, characterized by a box size of approximately a gigaparsec and a spatial resolution of roughly one kiloparsec. The extensive number of galaxy clusters generated within \hr\, coupled with its high resolution, empowers us to provide dependable insights into the generation and spatial distributions of ICL.

In \S2, we introduce our method and discuss its application.
In \S3, we define the simulated data we will use.
We describe the cluster components and dynamical state parameters used in the analysis in \S4. We present our results in \S5 and discuss the results in \S6. We conclude and summarize them in \S7.

\section{Weighted Overlap Coefficient Method} \label{sec:woc}
\begin{figure}
\begin{center}
\includegraphics[width=0.7\columnwidth,trim={1.5cm 0.5cm 1.5cm 5cm},clip]{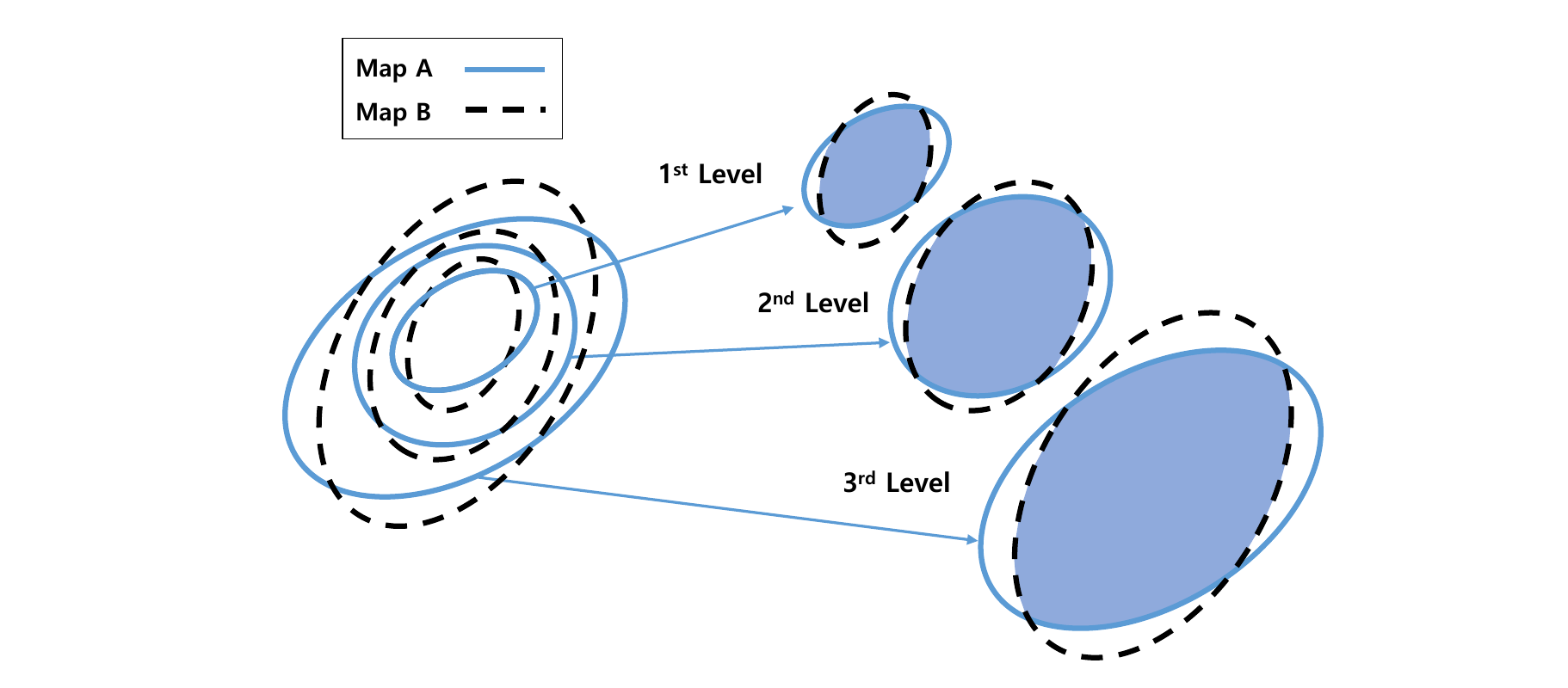}
\caption{Cartoon schematic of the WOC method.  It involves assessing the overlap of contours between two distributions at different density threshold levels. The fraction of this overlapping area is computed, and weights are applied based on signal strength to quantify their overall similarity, resulting in a normalized value ranging from 0 to 1. For additional information, refer to Section \ref{sec:woc}.
\label{fig:woc_idea}}
\end{center}
\end{figure}
We quantify the similarity between two distributions of surface brightness or density by using the \textit{Weighted Overlap Coefficient} \citep[WOC; ][]{2022ApJS..261...28Y}, which is a number between 0 and 1. Measurement of the WOC of two distributions proceeds as follows.
First, the two maps are smoothed over the same angular or spatial scale, 
and the areas enclosed by the iso-density contours are measured at a set of threshold levels.
Then, in the comparison map, we find the threshold levels of contours that enclose the same areas as those in the reference map. The degree of overlap of these matched areas between two distributions is used to calculate the WOC with our choice of weights that take into account contour area and threshold density.

Specifically, suppose we are comparing a reference map $A$ with a comparison map $B$ using $n$ sets of matched regions $A_i$ and $B_i$ with ${\rm area}(A_i)={\rm area}(B_i)$. Let the $i=1$ level correspond to the highest level. 
The WOC is defined as
\begin{equation}
\mathrm{WOC}(A,B)=\frac{ \sum\limits_{i=1}^{n} f_{i} \left(w_{i}+w_{\rho_A,i}+w_{\rho_B,i}\right)}{\sum\limits_{i=1}^{n} 
\left(w_{i}+w_{\rho_A,i}+w_{\rho_B,i}\right)},
\label{eq:woc}
\end{equation}
where $f_{i}={\rm area}(A_i \bigcap B_i) /{\rm area}(A_i)$ is the fraction of overlapping area between the $i$-th contours of two maps, $w_i={\rm area}(A_i)^{-1}/\sum_j {\rm area}(A_j)^{-1}$ is a normalized weight weighing higher-level contour areas more,  and 
$w_{\rho_A,i}=\rho_{A,i}/\sum_j \rho_{A,j}$ and $w_{\rho_B,i}=\rho_{B,i}/\sum_j \rho_{B,j}$ are normalized weights giving more weights to higher threshold values.
$\rho_{A,i}$ and $\rho_{B,i}$ denote the density threshold at $i^{th}$ level on map $A$ and map $B$, respectively.

Therefore, contour areas and threshold levels are taken into account in our weighting system.
The WOC parameter quantifies the spatial correspondence of two maps rather than their relative signal strengths or the exact shape of their profiles. The WOC method works well even for disconnected regions and does not require computation of individual contours, making it less biased when working with masked maps with boundaries.
Detailed information on the methodology, tests on robustness, and comparison with other methods can be found in \cite{2022ApJS..261...28Y}.

\section{Simulation Data} \label{sec:data}

\subsection{Horizon Run 5}
Horizon Run 5 (\hr) is a cosmological hydrodynamical zoomed-in simulation that aims to investigate galaxy formation and evolution in a cubic volume with a side length of $L_\mathrm{box} = 1049 ~\mathrm{cMpc}$. \hr\ {\bf is} run using a modified version \citep{2021ApJ...908...11L} of the adaptive mesh refinement code \texttt{RAMSES}~\citep{teyssier02}. \hr\ has a zoomed region of the cuboid geometry of $L_\mathrm{(x,y,z)}^\mathrm{zoom} = (1049, 119, 127 )$ cMpc crossing the central region of the simulation box. The cubic grids in \hr\ are refined down to $\Delta L\sim1\,$pkpc in the zoomed region. \hr\ adopts the cosmological parameters compatible with the results of \citet{planck16}: $\Omega_{\rm m}=0.3$, $\Omega_{\Lambda}=0.7$, $\Omega_{\rm b}=0.047$, $\sigma_8=0.816$. The initial condition of \hr\ is generated using the \texttt{MUSIC} package~\citep{hahn11} at $z=200$ based on the second-order Lagrangian perturbation theory~\citep[2LPT;][]{scoccimarro98,lhuillier14}. 

\hr\ has 256 (grid level 8) coarse grids on a side, and the zoomed region of a cuboid geometry initially has $8192\times930\times994$ grids at grid level 13. The high-resolution region is accordingly surrounded by padding grids of levels from 9 to 12. A dark matter particle has a mass of $6.89\times10^7\, M_{\odot}$ at level  13, and its mass rises by a factor of 8 with each decrease of grid level.  Grids are refined adaptively based on the octree mesh scheme down to $\Delta L\sim1\,$pkpc when their density exceeds the density of a grid enclosing eight dark matter particles at Level 13. \hr\ ends at the redshift of $z=0.625$ at which the age of the universe is $\sim7.7\,$Gyr for the cosmology adopted in \hr.

\texttt{RAMSES} has key physical ingredients in the form of subgrid recipes that govern the evolution of baryons. Gas cooling is computed using the cooling functions of \citet{sutherland93} in a temperature range of $10^4-10^{8.5}\,$K, and fine-structure line cooling is also implemented into \texttt{RAMSES} for a medium of the temperature down to $\sim750\,$K using the cooling functions of \citet{dalgarno72}. Cosmic reionization is approximated by assuming the assumption of a uniform UV background~\citep{haardt96}. Star formation rate is computed using the statistical approach of \citet{rasera06} based on the Schmidt law~\citep{schmidt59}. Supernova feedback operates in thermal and kinetic modes~\citep{dubois08}, and AGN feedback operates in radio-jet and quasar modes switched by the Eddington ratio~\citep{dubois12}. A massive black hole (MBH) initially has a mass of $10^4\, M_{\odot}$ and it is seeded in a grid when its gas density is higher than the threshold of star formation, and no other MBHs are found within 50 kpcs from the grid~\citep{dubois14b}. Gas accretion and MBH coalescence are the two mass growth channels of MBHs. The angular momentum of MBHs is computed by tracing the transfer of angular momentum from gas accretion and MBH coalescence~\citep{dubois14a}. The evolution of chemical abundance is computed using the methodology of \citet{few12} based on a Chabrier initial mass function~\citep{chabrier03}, and the abundance of H, O, and Fe are separately traced. Detailed information of \hr\ can be found in \cite{2021ApJ...908...11L}.

\begin{table}[t]
	\centering
        \caption{Specifications of the cosmological simulations for ICL studies.}
	\begin{tabular}{ll|ccccc|c}
            \hline Simulation & Type & $m_{\rm DM}$ & $m_{\star}$  & Spatial resolution\footnote{Minimum grid size for the AMR scheme-based codes, typical gravitational softening length for the SPH scheme-based codes.}  & Volume & Box size & Note \\
            &&[${\mathrm M_\odot}$]&[${\mathrm M_\odot}$]&[kpc]& [cMpc$^3$]&[cMpc]&\\
            \hline HR5 &    &  $6.9 \times 10^7$ & $2.5\times10^6$& 0.8 - 1.6 & $251.2^3$ \footnote{Effective volume of the initial zoomed-in region.}&1049&$\mu_{\mathrm {r}}^{\mathrm {lim}}\sim$ 29.5 mag/arcsec$^2$\footnote{M/L ratio is 2.95 in the MILES stellar population synthesis model package based on a Chabrier IMF and the BasTI isochrone.} \\
            \hline & TNG 300 & $5.9 \times 10^7$ & $1.1 \times 10^7$ & 1.0 - 2.0 & $302.6^3$ & 302.6 &\\
		  Illustris & TNG 100 & $7.5 \times 10^6$& $1.4 \times 10^6$ & 0.5 - 1.0 & $106.5^3$  & 110.7 &\\
             & TNG 50 & $4.5 \times 10^5$ & $8.1 \times 10^4$ & 0.195 - 0.39 & $51.7^3$  & 51.7 &\\
            \hline C-EAGLE & & $9.7 \times 10^6$ & $1.8 \times 10^6$ & 0.7 & &&$\mu_{\mathrm {lim}}\sim$ 30 mag/arcsec$^2$\\
            \hline GRT & & $5.4 \times 10^4$& & & &&$\mu_{\mathrm {lim}}\sim$ 32 mag/arcsec$^2$\\
            \hline Rudick +2006 & & $1.4 \times 10^6$ &  & & &&$\mu_{\mathrm {lim}}\sim$ 32 mag/arcsec$^2$\\
            \hline & Zoom 1& $8.3 \times 10^8$ & & 7.5 & &&\\
            Puchwein +2010 & Zoom 2& $1.1 \times 10^8$&  & 3.75 & &&ICL results converge  \\
             & Zoom 3 & $3.1 \times 10^7$&  & 2.5 & &&from Zoom 2\\
            & Zoom 4 & $1.3 \times 10^7$&  & 1.875 & &&\\
            \hline 
	\end{tabular}
	\label{tab:table1} 
\end{table}

\subsection{Structure Identification and Merger Trees}
\label{sec:structure_merger_trees}
The bounded objects and galaxies from the entire snapshot set of \hr\ are identified using the Physically Self-Bound (PSB)-based galaxy finder~\citep[\pgalf, see Appendix A of][for details]{kim23}. Firstly, Friends-of-Friends (FoF) objects are found from a unified data structure that contains all the mass components of dark matter, gas, stars, and MBHs by utilizing the adaptive FoF algorithm of \pgalf. Note that we replace the names FoF halos and subhalos with FoF objects and sub-objects, respectively, as we use all the mass components. The self-bound sub-objects are identified by measuring total energy and tidal radius from the local density peaks found in the coordinate-free stellar or dark matter density fields of FoF objects. 
Therefore, the sub-objects may be composed of dark matter or stellar mass only or multiple components. A sub-object is defined as a galaxy in \hr\ when gravitationally bound stellar mass larger than $10^7\,M_\odot$ is contained. The mass components that have positive total energy relative to the density peaks of any sub-objects within an FoF object are classified as unbound components. Specifically, the unbound stellar particles in a cluster-scale object are defined as ICL.

The merger trees of self-bound sub-objects are constructed by tracing stellar particles for galaxies and the most bound matter particles (MBPs) for the objects that do not contain stellar particles. An MBP is defined as a dark matter particle located at the deepest potential of a self-bound subhalo~\citep{hong16}. It is assumed that the motion of an MBP follows that of its subhalo. The advantage of the MBP scheme is that a broken branch in a merger tree can be repaired by tracking the MBP, which never disappears in a simulation box. Since the merger trees of \hr\ are constructed based on sub-objects, the evolution history of a FoF object is traced using the merger tree of its most massive sub-object, or in other words, its central galaxy. Further details of the tree-building algorithm of \hr\ can be found in \citet{park22} and \citet{lee23}.

In the last snapshot of \hr\ ($z=0.625$), there are 63 clusters with high FoF mass ($M_{\rm tot}>  10^{14} M_{\odot}$) in the zoomed region. The age of the universe is 7.7 Gyr at $z=0.625$, and thus, there should be some relatively less-massive objects that would form cluster-scale objects by $z=0$. From \texttt{HR5-Low}, a low-resolution simulation of \hr, it is inferred that massive objects typically double their total mass between $z=0.625$ and 0 \citep{lee23}. Consequently, we identified 174 objects with $M_{\rm tot}> 5\times10^{13} M_{\odot}$ at $z=0.625$, which later reach $M_{\rm tot}> 1\times10^{14} M_{\odot}$ at $z=0$ . These objects are located within the zoomed region of \hr\ and remain uncontaminated by low-level particles. This serves as the primary cluster sample for this study.

\subsection{ICL in \hr}
 
As we are interested in ICL, a low-surface brightness unbound component in clusters, the resolution of the simulation can be of concern.  
A higher resolution enables one to explore the ICL at lower surface brightness but inevitably reduces the simulation volume, which results in reduced and biased statistics due to fewer structures and missing large-scale fluctuations. 
In Table~\ref{tab:table1}, we list the simulation parameters of a few cosmological simulations used for previous ICL studies together with those of \hr.
Table~\ref{tab:table1} shows that \hr\ has a spatial resolution between those of Illustris TNG100 and TNG300 \citep{2018MNRAS.475..624N, 2018MNRAS.473.4077P}; its dark matter mass resolution is close to that of TNG300, while its stellar mass resolution is comparable to that of TNG100. On the other hand, the simulation box sizes of both TNG300 and TNG100 are significantly smaller than that of \hr. 
TNG100 and TNG50 have higher resolutions than \hr, but they do not have a sufficient number of simulated galaxy clusters (10 in TNG100, only a few in TNG50), let alone the missing large-scale power.

\cite{2010MNRAS.406..936P} examined the amount of ICL in a suite of zoomed simulations with various resolutions and showed that the amount of ICL starts to converge at the spatial resolutions below $\sim4\,$pkpc. Therefore, we conclude that \hr\ is suitable for the study of ICL due to its large simulation box size for capturing large-scale power, large cluster sample size giving good statistics, and mass and spatial resolutions sufficient for ICL formation simulation. A caveat of this study is that the ICL components are defined to be unbound stellar particles in FoF objects. This definition may not be fully consistent with observations since, in practice, it is difficult to separate the light of unbound stars from the light of all stars in an observed cluster.

\section{Analysis} \label{sec:analysis}
\subsection{Candidates for Dark Matter Tracer} \label{subsec:comp}

\begin{figure}
\centering
\includegraphics[width=0.95\textwidth,trim={3cm 1cm 3cm 1.5cm},clip]{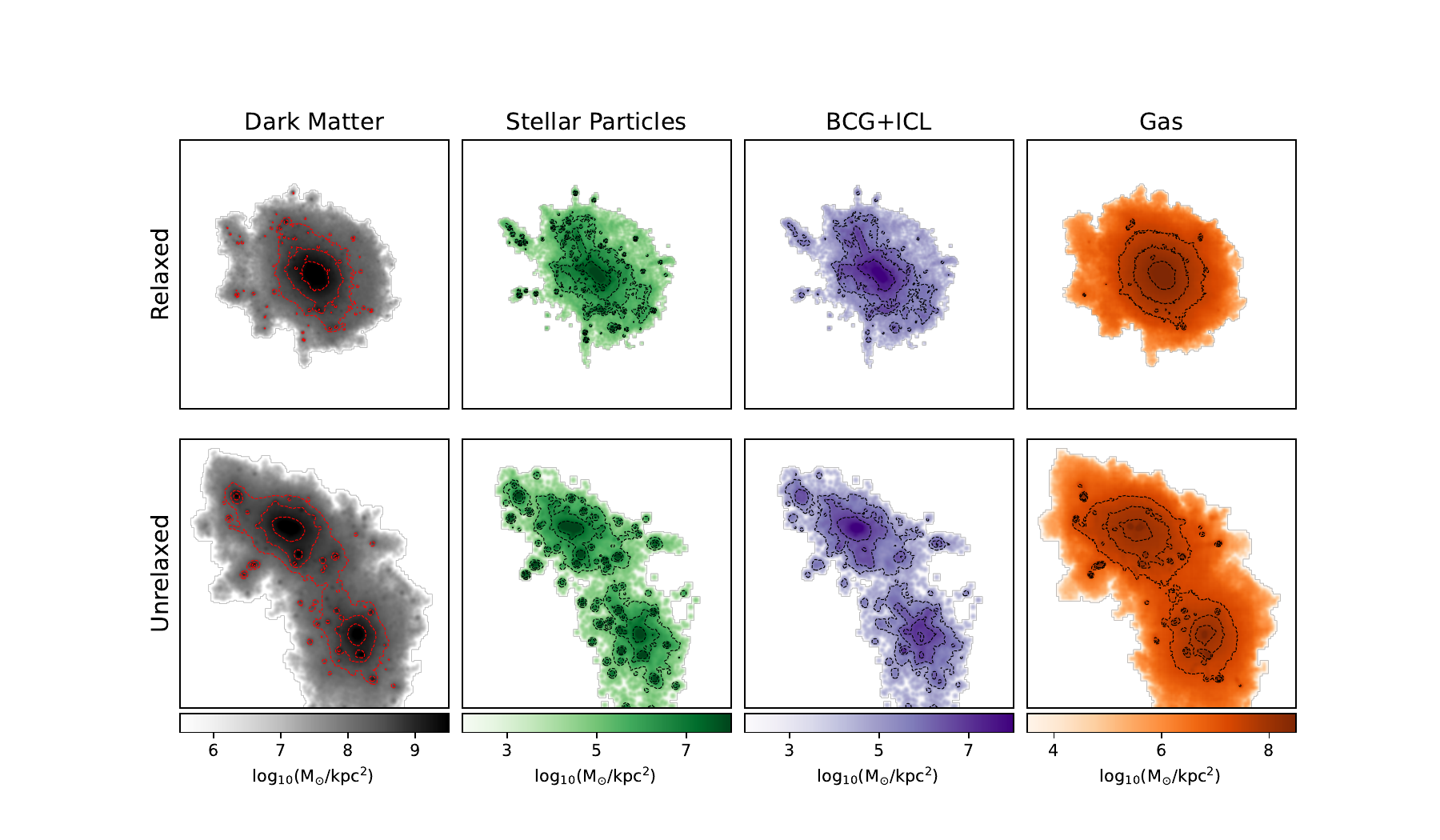}
\caption{Projected views of various components of a typical relaxed (upper panel) and unrelaxed (lower panel) galaxy clusters in \hr\ simulation.
The scales of images are 2 $r_{\text{vir}} \times$ 2 $r_{\text{vir}}$ of the cluster, where the virial radius is 1.39 cMpc for the relaxed and 0.82 cMpc for the unrelaxed cluster.  
For each galaxy cluster, the surface density distributions of dark matter, all stars (BCG+ICL+satellite galaxies), BCG+ICL, and gas components are shown. Dashed lines denote the density levels of azimuthally-averaged radial density profile at the 0.1, 0.2, and 0.3 virial radii, for which the WOC, our similarity measure between components, is measured (see details in Section \ref{subsec:result1}).}
\label{fig:example2}
\end{figure}

Which component of the galaxy cluster best matches the spatial distribution of dark matter? 
To answer the question, we consider four mass components or a combination of components as follows.

 {\bf Stellar particles:} 
 In this study, the stellar particle component includes all stellar particles within the brightest cluster galaxy (BCG), satellite galaxies, and ICL. We visualize these elements in the second column of Figure \ref{fig:example2}, depicting a relaxed and an unrelaxed galaxy cluster in the \hr\ simulation. 
 This component can be most straightforwardly defined in both simulations and observations, as it does not require distinguishing between individual galaxies and the smooth, diffuse ICL.

{\bf Galaxies:} In this study, `galaxies' refers to stellar particles that are gravitationally bound to the galaxies cataloged in the \hr~dataset. 
Our approach does not set a specific mass threshold, thus including even dwarf galaxies with a FoF mass around $\sim 10^7 M_{\odot}$. While these smaller galaxies present observational challenges due to their diffuse nature, we consider them as part of our comprehensive analysis.
Galaxies, particularly the more massive ones, are often the most accessible probes for locating dark matter in observational astronomy. Excluding diffuse dwarf elliptical galaxies, most cluster member galaxies are relatively straightforward to observe, even without deep imaging techniques.
Moreover, satellite galaxies within these clusters can contain significant amounts of dark matter, with their contribution varying based on their mass or luminosity. 
Previous studies, such as those involving dense redshift surveys, have demonstrated the potential of galaxy number density as a proxy for estimating the total mass distribution within clusters \citep{2014ApJ...797..106H}. Additionally, a recent simulation-based study advocated using a galaxy mass-weighted number density distribution as a dark matter tracer \citep{2022ApJ...934...43S}. This approach aligns with our aim to explore various components within galaxy clusters as potential dark matter proxies.

{\bf BCG+ICL:} In this study, the BCG is defined as the ensemble of stellar particles gravitationally bound to the most massive galaxy within the cluster. In observational terms, the BCG is commonly regarded as the `brightest' cluster galaxy. Therefore, we designate the most massive galaxy in terms of stellar mass as the BCG for the purposes of this study.
The ICL is the stellar particles that are gravitationally bound to the cluster itself but not to any specific galaxy within it. 
The BCG is typically located at the deepest point of the gravitational potential of the cluster, while the ICL is broadly distributed on cluster scales \citep{1998ApJ...502..141D}.
Studies have shown that the velocity dispersion of ICL mirrors that of the cluster, suggesting a close following of the gravitational potential of the cluster \citep{2010MNRAS.406..936P, 2020MNRAS.491.2617E, 2020ApJ...894...32G, 2021MNRAS.500.3462M}. The BCG and ICL as a combined component are depicted in the third column of Figure \ref{fig:example2}, showing their distribution in both relaxed and unrelaxed clusters in the \hr\ simulation. This combined BCG+ICL component has been increasingly recognized as an effective tracer for dark matter \citep{2019MNRAS.482.2838M, 2020MNRAS.494.1859A, 2022ApJS..261...28Y}. Its observational advantage lies in its accessibility through optical deep imaging, which is often simpler compared to the techniques required for gas component analysis via X-ray or radio observations.

{\bf Gas:} In this study, the gas component encompasses all gas particles within the cluster, whether they are gravitationally bound to individual galaxies or not.
The gas component in galaxy cluster occupies around $\sim$ 12\% of cluster mass, whereas $\sim$ 85\% for dark matter and only $\sim$ 3\% for stars \citep{2009ApJ...693.1142S,2013ApJ...778...14G, 2013MNRAS.429.3288S}.
Traditionally, the X-ray observed hot gas representing the gas trapped within the gravitational potential well has been used to trace the total mass of galaxy cluster  \citep{2001Natur.409...39B}. 
 In our study, visualizations of the gas component within the \hr\ galaxy clusters can be seen in the fourth column of Figure \ref{fig:example2}. These images highlight the gas distribution in relaxed and unrelaxed clusters, providing a comparative view of the other components.

\subsection{Indicators for Dynamical State} \label{subsec:dynamic}
\begin{figure}
\centering
\includegraphics[width=0.8\textwidth,trim={0 0 0 1cm},clip]{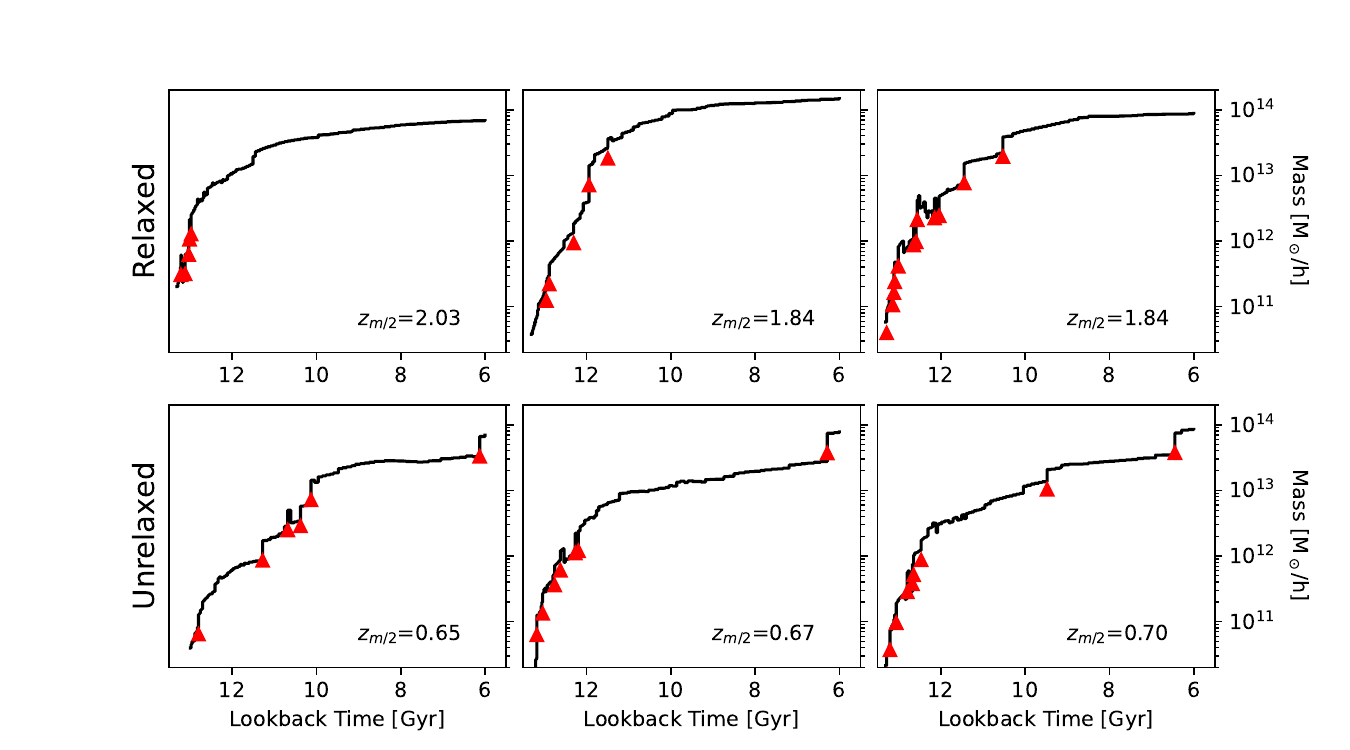}
\caption{Cumulative mass growth histories of representative galaxy clusters within the \hr\ simulation. Red triangles denote major merger events (mass ratio 1:3 or less). The redshift at which clusters achieve half of their final mass ($z_{m/2}$) is provided. (Upper row) Depicts relaxed galaxy cluster cases, showing early major mergers and an extended absence of significant merging events, resulting in an early half-mass epoch. (Lower row) Highlights unrelaxed galaxy cluster cases characterized by more recent major merging events and a later half-mass epoch.}
\label{fig:massgrowth}
\end{figure}

Several indicators for the dynamical state of galaxy clusters have been suggested in previous studies. We will inspect the correlation of our WOC between cluster components with these dynamical state indicators listed below.

{\bf Half-mass epoch ($z_{m/2}$):} 
The concept of the half-mass epoch is pivotal in understanding the dynamical evolution of galaxy clusters within the framework of the $\Lambda$CDM cosmology. Galaxy clusters grow through hierarchical mergers, and the half-mass epoch — denoted as $z_{m/2}$ — provides a key temporal marker in this process. Figure \ref{fig:massgrowth} illustrates the mass growth histories of selected relaxed and unrelaxed clusters in the \hr\ simulation, with major mergers (mass ratios of 1:3 or less) highlighted by red triangles.
In this study, $z_{m/2}$  refers to the redshift at which a cluster accrues half of its final observed mass. Typically, relaxed clusters exhibit a high $z_{m/2}$, indicating that they have achieved significant mass at an earlier epoch, followed by a period of relative stability with fewer major mergers. Conversely, unrelaxed clusters, characterized by a lower $z_{m/2}$, have undergone substantial mass accumulation in more recent epochs, often through recent major merging events.
By analyzing the mass growth histories of the 174 galaxy clusters in the \hr\ simulation, we determine the $z_{m/2}$ for each cluster. This measurement is a critical indicator of the dynamical state of the cluster, as it encapsulates the cumulative effect of its mass growth history. The differentiation between relaxed and unrelaxed systems based on $z_{m/2}$ is supported by various studies \citep{1996MNRAS.281..716C, 2021A&A...651A..56G, 2022ApJS..261...28Y, 2023ApJ...943..148C}, reinforcing its validity as a robust metric in our analysis.

{\bf Magnitude gap ($\Delta M_{12}$):}
The magnitude gap, represented as $\Delta M_{12}$, is a significant measure for assessing the dynamical state of galaxy clusters \citep{1994Natur.369..462P, 2005ApJ...630L.109D, 2006AJ....132..514C}. It is defined as the difference in absolute magnitude between the BCG and the second brightest galaxy in the cluster. This metric offers insights into the evolutionary history of the cluster, particularly in the context of hierarchical merging processes.
Clusters classified as `fossil clusters' are a prime example of systems where $\Delta M_{12}$ is notably large, typically exceeding two magnitudes in the \textit{r}-band considering member galaxies within 0.5 $r_{\text{vir}}$ of the  cluster. These clusters, also characterized by extended X-ray emissions with luminosities greater than $10^{42}erg/s$, are generally considered to be in a more relaxed state \citep{2003MNRAS.343..627J}. The substantial magnitude gap in these clusters often indicates a history of significant merging events, resulting in the dominance of the BCG and a paucity of comparably bright galaxies.
In observational studies, $\Delta M_{12}$ is particularly valued for its relative ease of measurement. By focusing on the two brightest galaxies within a cluster, we can quickly glean information about the dynamical state of the cluster without the need for extensive and complex data analysis. This simplicity makes $\Delta M_{12}$ an effective and accessible tool for evaluating the evolutionary stage of galaxy clusters.
However, it has been suggested that the magnitude gap does not genuinely reflect the final relaxation, but rather the fossil configuration is just a transitional state \citep{2022ApJ...928..170K,2022ApJ...936...59D}. Considering this we chose not to apply the 0.5 $r_{\text{vir}}$ cut, i.e., we considered all member galaxies in the calculation of the magnitude gap in an effort to increase the robustness of the magnitude gap as a dynamical state indicator. Despite this, the inherent limitations of $\Delta M_{12}$ as a precise measure of dynamical relaxation remain, leading us to adopt it primarily for its convenience.

{\bf Offset:} 
The offset metric serves as an indicator of the dynamical state of galaxy clusters, particularly in the context of their merging history. This measure evaluates the spatial discrepancy between the center of mass of the cluster and the position of its most significant density peak, typically represented by the central galaxy.
In actively merging clusters, the internal structure is often disturbed, leading to a notable offset. This displacement is indicative of ongoing dynamical processes, such as merging events, which disrupt the equilibrium of the gravitational potential of the cluster. In observational terms, the center of mass can be inferred from the distribution of dark matter, as revealed through weak-lensing analysis, or from the distribution of hot gas, as observed in X-ray emissions. The density peak, on the other hand, is commonly associated with the location of the BCG.
In our analysis of the \hr\ simulation, we quantify the offset by measuring the distance between the center of the cluster and the BCG. This distance is normalized by the virial radius ($R_{200}$) of the cluster, providing a standardized measure of offset across different clusters. The use of this metric allows us to gauge the relaxedness of a system, with a smaller offset suggesting a more relaxed state \citep{2017MNRAS.464.2502C, 2021A&A...651A..56G}, where the mass of the cluster and its gravitational center are closely aligned.

{\bf ICL fraction ($f_\mathrm{ICL}$):} 
The fraction of ICL within a galaxy cluster offers vital insights into the dynamical maturity and history of the cluster. The abundance of ICL increases through successive galaxy interactions in galaxy clusters, as shown in simulation studies \citep{2007MNRAS.377....2M, 2007ApJ...666...20P, 2007ApJ...668..826C, 2010MNRAS.406..936P, 2011ApJ...732...48R, 2014MNRAS.437.3787C, 2015MNRAS.451.2703C}. Observational studies also report several scenarios for the production of intracluster stars, including BCG major mergers followed by violent relaxation \citep{2007ApJ...665L...9R, 2018ApJ...862...95K, 2023Natur.613...37J}, tidal stripping from the outskirts of L$^{\ast}$ member galaxies \citep{2017ApJ...851...75I, 2018MNRAS.474.3009D, 2018MNRAS.474..917M}, disruptions of dwarf galaxies as they fall towards the center of the galaxy cluster \citep{2011MNRAS.414..602T} and in-situ star formation \citep{2002ApJ...580L.121G}. 
Thus, we expect the more mature galaxy clusters with interaction-rich history to have higher ICL amounts. 
In the meantime, \citet{2023Natur.613...37J} showed that the ICL fraction is independent of the cluster mass.  Moreover, there is observational evidence supporting that the ICL fraction measured at certain wavelengths increases during merging events, i.e., the unrelaxed system could show higher ICL fraction \citep{2018ApJ...857...79J, 2019A&A...622A.183J, 2021ApJ...922..268J, 2023A&A...676A..39J, 2022MNRAS.512.1916D}.

{\bf BCG+ICL fraction ($f_\mathrm{BCG+ICL}$):} 
Significant dynamic events within the cluster, such as mergers and tidal interactions, contribute to the accumulation of mass in both the BCG and ICL \citep{2014MNRAS.437.3787C, 2019ApJ...871...24C}. 
We expect that more relaxed systems have had more time to have dynamical interactions and that that would result in more accumulative BCG and ICL.
This results in an increased BCG+ICL fraction, making it a valuable probe of the evolutionary history of the cluster. 
The BCG+ICL fraction has been explored by numerous observational \citep{2004ApJ...617..879L, 2005MNRAS.358..949Z, 2011MNRAS.414..602T, 2013ApJ...778...14G, 2014A&A...565A.126P} and simulation studies \citep{2010MNRAS.406..936P, 2015MNRAS.451.2703C, 2017MNRAS.467.4501H, 2018MNRAS.475..648P, 2022ApJS..261...28Y, 2023ApJ...943..148C}. 
In this study, both the ICL fraction and BCG+ICL fraction are measured in stellar mass.

{\bf Substructure fraction ($f_\mathrm{sub}$):}
The substructure fraction within a galaxy cluster, denoted as $f_\mathrm{sub}$, is a critical indicator of its dynamical state.
In more relaxed clusters, we typically observe a dominant BCG and smaller substructures, leading to a lower $f_\mathrm{sub}$. Conversely, unrelaxed clusters, often in the throes of active merging and accretion, show a higher $f_\mathrm{sub}$, indicating a greater presence and influence of substructures. In this study, $f_\mathrm{sub}$ is calculated by summing the stellar masses of satellite galaxies and normalizing this sum by the total stellar mass of the cluster, i.e., $f_\mathrm{sub}=\sum M_\mathrm{sub}/M_\mathrm{tot}$, where $M_\mathrm{sub}$ is the mass of each substructure without taking the mass of the most massive substructure into account, and $M_\mathrm{tot}$ is the total mass of the system. This leads to $f_\mathrm{sub}=1-f_\mathrm{BCG + ICL}$.
This approach effectively mirrors the internal mass distribution of the cluster and provides insights into its current dynamical state \citep{2017MNRAS.464.2502C, 2021A&A...651A..56G}.

\begin{figure}
\centering
\includegraphics[width=0.95\textwidth,trim={3cm 0 3cm 1cm},clip]{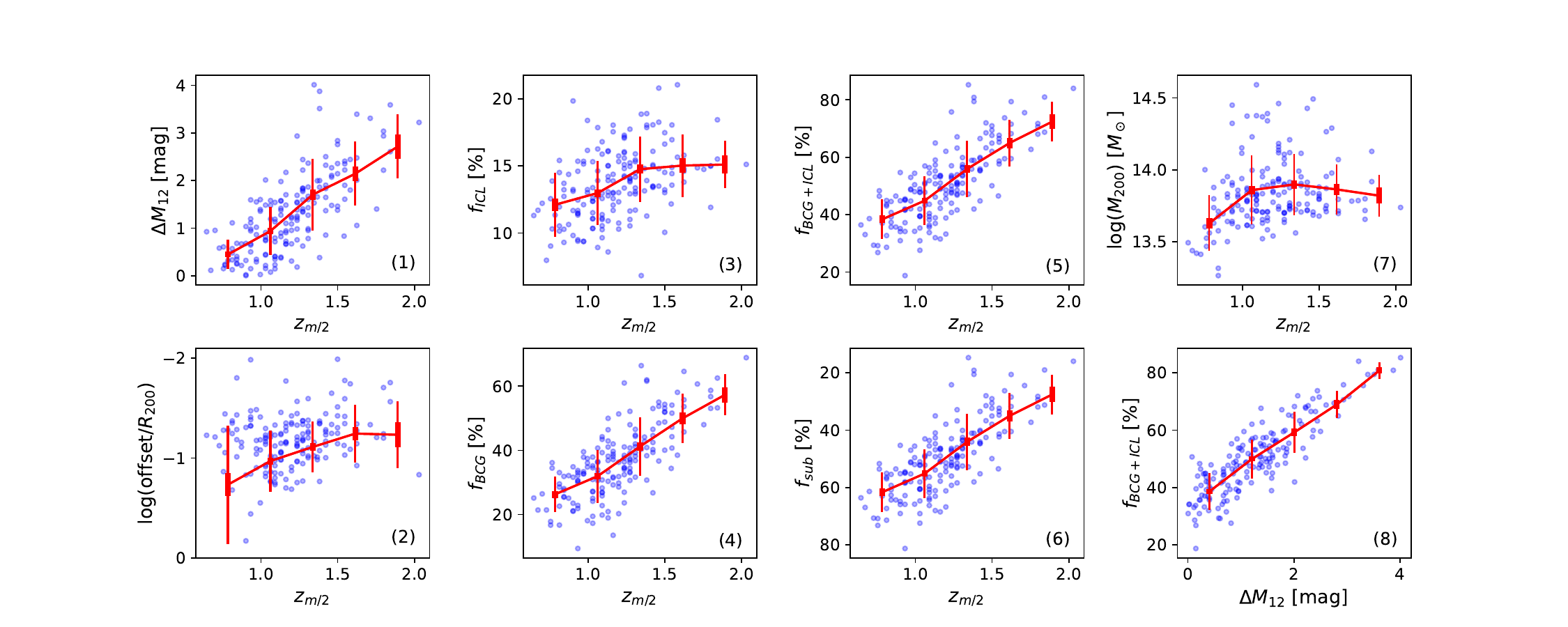}
\caption{Various dynamical state parameters used in this study and the relation between the parameters. (Top) The magnitude gap between BCG and the second brightest galaxy, ICL fraction, BCG+ICL fraction, and $M_{200}$ for different half-mass epochs are plotted. (Bottom) BCG - cluster center offset, BCG fraction, substructure fraction for different half-mass epochs, and BCG+ICL fraction for different magnitude gaps are plotted. Thin vertical lines are standard deviations, and thick vertical lines are standard errors. For every sub-plot, increasing the $x$-axis and increasing the $y$-axis is supposed to indicate a more relaxed state. The $M_{200}$ (panel nr.7) does not necessarily represent the dynamic state but is included for completeness.}
\label{fig:dyn}
\end{figure}

The virial ratios ($\eta$ = 2$T/|W|$, where $T$ and $W$ are the kinetic energy and the gravitational potential energy, respectively) would be the most theoretical indicator of the virialization of the system. However, the calculation of the gravitational potential of each particle is computationally expensive. Previous studies calculate the virial ratio through random sampling, sometimes adding a surface pressure term \citep{2022ApJ...934...43S}, which causes possible noise. Moreover, for each particle, the gravitational potential would be influenced by all the other particles, including the one not belonging to the cluster. We could not take into account these particles in the calculation, which will make the calculated virial ratio inaccurate. Therefore, we decided not to use the virial ratio parameter in this study. 

Looking at the relation between dynamical indicators (see Figure \ref{fig:dyn}), the half-mass epoch ($z_{m/2}$),  the magnitude gap between BCG and the second brightest galaxy ($\Delta M_{12}$), and the BCG+ICL fraction (substructure fraction) appear to be good (stable) indicators.
While the $M_{200}$, the offset between cluster center and BCG, and the ICL fraction show less correlation with other indicators, it is remarkable that the BCG+ICL fraction has a tight relation with the magnitude gap parameter with the least outlier (panel nr.8 in Figure \ref{fig:dyn}). While the half-mass epoch ($z_{m/2}$),  is a useful dynamical state indicator of a galaxy cluster, one may posit that a combination of indicators may provide a more robust estimate of this quantity \citep{hyowon}. However, since this is an ongoing area of investigation, we chose to simply use $z_{m/2}$ alone as the dynamical state indicator.

\section{Results} \label{sec:result}
\subsection{Similarity of spatial distributions}
\label{subsec:result1}
In this section, we explore the efficacy of different galaxy cluster components in tracing dark matter distributions, using the WOC as our primary tool of analysis. The components under consideration include (1) all stellar particles (comprising BCG, satellite galaxies, and ICL), (2) stellar particles of just the cluster galaxies (BCG and satellite galaxies), (3) stellar particles of BCG and ICL, and (4) all gas particles.

The WOC calculations were conducted using bins set at 0.1, 0.2, and 0.3 times the virial radii ($R_{200}$) of each cluster. In this analysis, dark matter serves as the reference map against which the spatial distributions of stars (including BCG, satellite galaxies, and ICL), galaxies (BCG and satellite galaxies), BCG+ICL, and gas are compared. Notably, the center for each radial profile within the WOC calculation is aligned with the peak of the dark matter density. This approach allows for a consistent comparison, as it does not require setting a common center in the comparison maps since the contour density levels of equal area are independently computed for each map.

\begin{table}[t]
	\centering
 	\caption{WOC results summary. Among 174 clusters, 55 early formed ($z_{m/2}\geq 1.307$) clusters are grouped as Relaxed, 54 late formed ($0 < z_{m/2}\leq 1.029$) clusters as Unrelaxed, and in between ones as Middle. There are two galaxy clusters where the half-mass epoch ($z_{m/2}$) is not measured, which are excluded from the grouping. WOC results of each group are calculated by taking the median ($\pm$ standard deviation) of WOC results of the individual clusters, for all three projections ($xy, yz, zx$). The median standard deviations of the WOC result of three different projections are indicated below the WOC value. The number of clusters with all three projection measurements is indicated below. }
	\begin{tabular}{c|cccc}
            \hline & All 174 clusters & Relaxed 55 clusters & Middle 63 clusters & Unrelaxed 54 clusters \\
            \hline WOC (DM, star) & 0.661 ($\pm$ 0.051)& 0.662 ($\pm$ 0.049)& 0.658 ($\pm$ 0.054)& 0.661 ($\pm$ 0.047)\\
            std ($xy, yz, zx$)  & 0.018 & 0.017 & 0.019 & 0.017 \\
            Nr. of measurement & 172 & 55 & 62 & 53 \\
            \hline WOC (DM, galaxy) &  0.557 ($\pm$ 0.047)& 0.545 ($\pm$ 0.036)& 0.552 ($\pm$ 0.040)& 0.587 ($\pm$ 0.057)\\
            std ($xy, yz, zx$) &  0.021 & 0.020 & 0.020 & 0.022 \\
            Nr. of measurement & 80 & 21 & 30 & 28 \\
            \hline WOC (DM, BCG+ICL) & 0.792 ($\pm$ 0.049)& 0.826 ($\pm$ 0.035)& 0.784 ($\pm$ 0.042)& 0.768 ($\pm$ 0.050)\\
            std ($xy, yz, zx$) & 0.024 & 0.015 & 0.025 & 0.039 \\
            Nr. of measurement & 172 & 55 & 62 & 53 \\
            \hline WOC (DM, gas) & 0.813 ($\pm$ 0.044)& 0.835 ($\pm$ 0.036)& 0.800 ($\pm$ 0.042)& 0.791 ($\pm$ 0.042)\\
            std ($xy, yz, zx$) & 0.024 & 0.018 & 0.027 & 0.025 \\
            Nr. of measurement & 174 & 55 & 63 & 54 \\
            \hline WOC (DM, BCG+ICL+gas)\footnote{see Section \ref{sec:discusion}} & 0.862 ($\pm$ 0.028)& 0.887 ($\pm$ 0.024)& 0.851 ($\pm$ 0.022)& 0.853 ($\pm$ 0.023)\\
            std ($xy, yz, zx$) & 0.019 & 0.014 & 0.022 & 0.022 \\
            Nr. of measurement & 174 & 55 & 63 & 54 \\
            \hline 
	\end{tabular}

	\label{tab:table2} 
\end{table}

\begin{figure}
\centering
\includegraphics[width=1.0\textwidth,trim={4cm 1cm 4cm 1cm},clip]{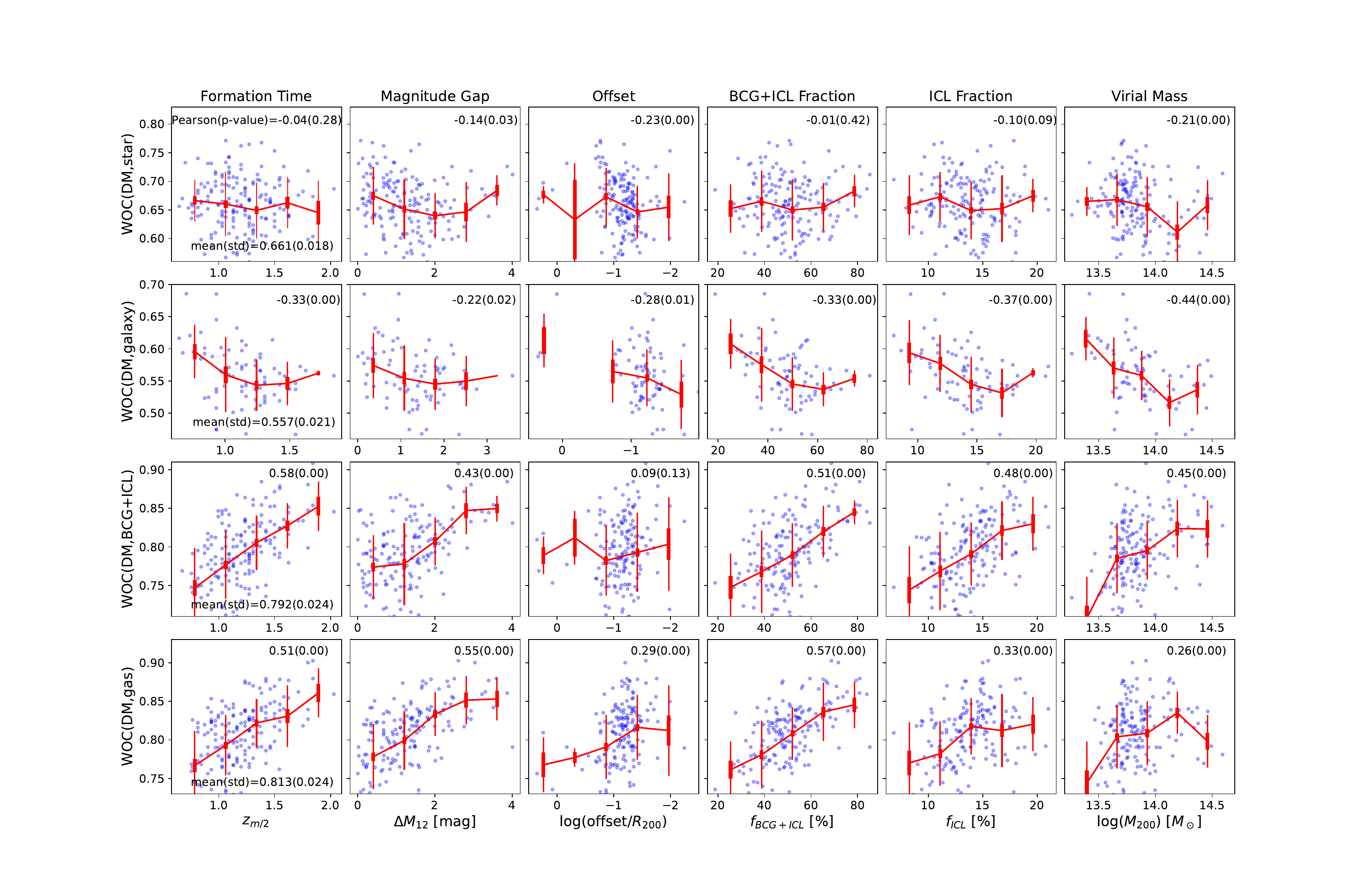}
\caption{WOC results for various components and dynamical state indicators. Spatial distribution similarity between dark matter and all stellar particles (first row), all galaxies (second row), BCG+ICL (third row), and gas particles (fourth row) are plotted for various relaxedness proxies. For every subplot, increasing the $x$-axis represents a more relaxed state. Blue dots are the WOC results for individual galaxy clusters. Red trend lines represent the median value of WOC in each bin, thin vertical red lines are standard deviations, and thick vertical red lines are standard errors. Pearson coefficients and p-values are indicated in the upper right corner of each subplot.}
\label{fig:woc_dyn}
\end{figure}

{\bf Luminous tracer for dark matter}: 
Our primary objective was to identify which cluster component most accurately traces the spatial distribution of dark matter. This was achieved by calculating the WOC, at z=0.625, between dark matter and the aforementioned components across all 174 galaxy clusters included in our study (refer to Table \ref{tab:table2}). For each cluster, we averaged results from three different projection angles ($x-y$, $x-z$, and $y-z$) and subsequently computed the median of these averages for the entire cluster set. The standard deviation of the WOC values across these projection angles was also determined (see std ($xy, yz, zx$) in Table \ref{tab:table2}).

It is worth noting that for the galaxies component, due to its high concentration in the cluster centers, WOC values compared to dark matter could not be measured for a significant portion of our sample. The problem arises in the algorithm because it must choose a contour level in map2 (e.g galaxies) that has an equal area as a contour in the reference map1 (DM). However, the galaxy field being much more spatially concentrated compared to DM means that sometimes an equal area having contour cannot be found \citep[see][for further details]{2022ApJS..261...28Y}. Out of 174 clusters, WOC calculations for the dark matter versus galaxies comparison were successful in all three projection angles only for 80 clusters. The results presented in Table \ref{tab:table2} pertain only for the cases where WOC was measurable in all three projections. An attempt to include clusters with one or two projection angles yielded similar results, reaffirming the robustness of our findings.

The findings reveal distinct variations in how well different components trace dark matter. Among the analyzed components, the galaxies map showed a notably lower WOC with dark matter (average WOC (DM, galaxies) = 0.557), indicating a less similar spatial distribution than other components. 
In contrast, both the BCG+ICL and gas components exhibited a higher degree of similarity with the dark matter distribution, with WOC (DM, BCG+ICL) and WOC (DM, gas) scoring 0.792 and 0.813, respectively. This suggests that these components, particularly gas and the combination of BCG and ICL, are more effective tracers of dark matter within galaxy clusters.
This result is consistent with the previous GRT simulation result \citep{2022ApJS..261...28Y}, which showed that the BCG+ICL trace dark matter more faithfully than the star (BCG+ICL+satellite galaxies). 

This difference in tracing efficacy can be attributed to distinct dynamical origins. 
For instance, satellite galaxies that have recently entered the cluster from various directions along filaments introduce a higher degree of anisotropy, compared to more uniformly distributed dark matter. Conversely, the BCG  often are an elliptical galaxy with a smooth, extended envelope, while the ICL, characterized by higher velocity dispersion, tends to have a round and extended shape. This morphological similarity with the smoother gravitational potential shape of the dark matter enhances the tracing accuracy of the BCG+ICL component.
Galaxies experience much stronger dynamical friction than the ICL, as they sink into the gravitational potential of the galaxy cluster.
This friction alters their orbits, leading to increased deviations from the mean-field defined by dark matter. In contrast, the collisionless ICL undergoes secular relaxation over time, gradually aligning its distribution more closely with that of the dark matter. Collectively, these factors elucidate why BCG+ICL emerges as a more faithful tracer of dark matter compared to the composite star component.

Additionally, the WOC analysis across different projection angles revealed that the WOC (DM, star) exhibited a minor standard deviation (0.018), suggesting a consistent spatial correlation irrespective of the viewing angle. However, the WOC (DM, BCG+ICL) and WOC (DM, gas) displayed slightly larger deviations (0.024), indicating some variation in similarity depending on the projection.  

{\bf Probing the evolutionary state of galaxy clusters}: 
We explored the potential of spatial distribution similarities between dark matter and various cluster components as probes of dynamical states of the galaxy cluster. Specifically, we examined whether the degree of alignment between these distributions correlates with the level of relaxedness in a cluster. This analysis involved assessing the WOC between dark matter and different components (stars, galaxies, BCG+ICL, and gas) against various dynamical state markers established in Section \ref{subsec:dynamic}, including the half-mass epoch ($z_{m/2}$), magnitude gap ($\Delta M_{12}$), BCG-cluster center offset, and the fractions of BCG+ICL and ICL (illustrated in Figure \ref{fig:woc_dyn}).

Our findings, as depicted in Figure \ref{fig:woc_dyn}, reveal a clear relationship between the WOC values and the dynamical states of clusters. Higher WOC values, indicating greater similarity in spatial distribution with dark matter, were observed in more relaxed clusters. This trend was particularly pronounced for the BCG+ICL and gas components, suggesting these elements are more reliable tracers of dark matter in dynamically mature clusters. Conversely, the WOC for stars or galaxies did not show a significant trend correlating with increased relaxation of the clusters.

A notable pattern emerged when examining the WOC in relation to the half-mass epoch ($z_{m/2}$). The correlation strength, quantified using the Pearson coefficient, was highest for WOC (DM, BCG+ICL) at 0.58, suggesting a strong relationship between the dynamical maturity of a cluster and the spatial alignment of its BCG+ICL component with dark matter. Similarly, WOC (DM, gas) demonstrated a high correlation with the BCG+ICL fraction at 0.57, further underscoring the relevance of these components in tracing dark matter in more evolved galaxy clusters. 

These correlations are not only statistically significant but also offer practical insights. For instance, in clusters with relatively high BCG+ICL fractions (60–80\%), we can trace approximately 80–85\% of dark matter using the BCG+ICL distribution.  
Similarly, in the system with a high BCG+ICL fraction, we could trace dark matter around 82-84\% using the gas distribution.
These results highlight the potential of using specific cluster components as proxies for dark matter distribution, particularly in assessing the dynamical states of galaxy clusters.

The trends we find in the data can be clearly seen considering the standard error. However, for the application to individual galaxy clusters, we should be cautious, considering the much larger standard deviation.

\subsection{WOC Evolution} \label{subsec:evolution} 
In this part of our study, we focused on how the similarity in spatial distribution between dark matter and key cluster components (BCG+ICL and gas) evolves across different epochs. Utilizing the WOC method, we traced the evolution of these similarities from various redshifts to the final snapshot at $z=0.625$, as outlined in Section \ref{sec:structure_merger_trees}. This involved tracking the progenitors of the BCGs using merger trees and measuring the WOC for the host systems at each epoch.

\begin{figure}
\centering
\includegraphics[width=0.9\textwidth]{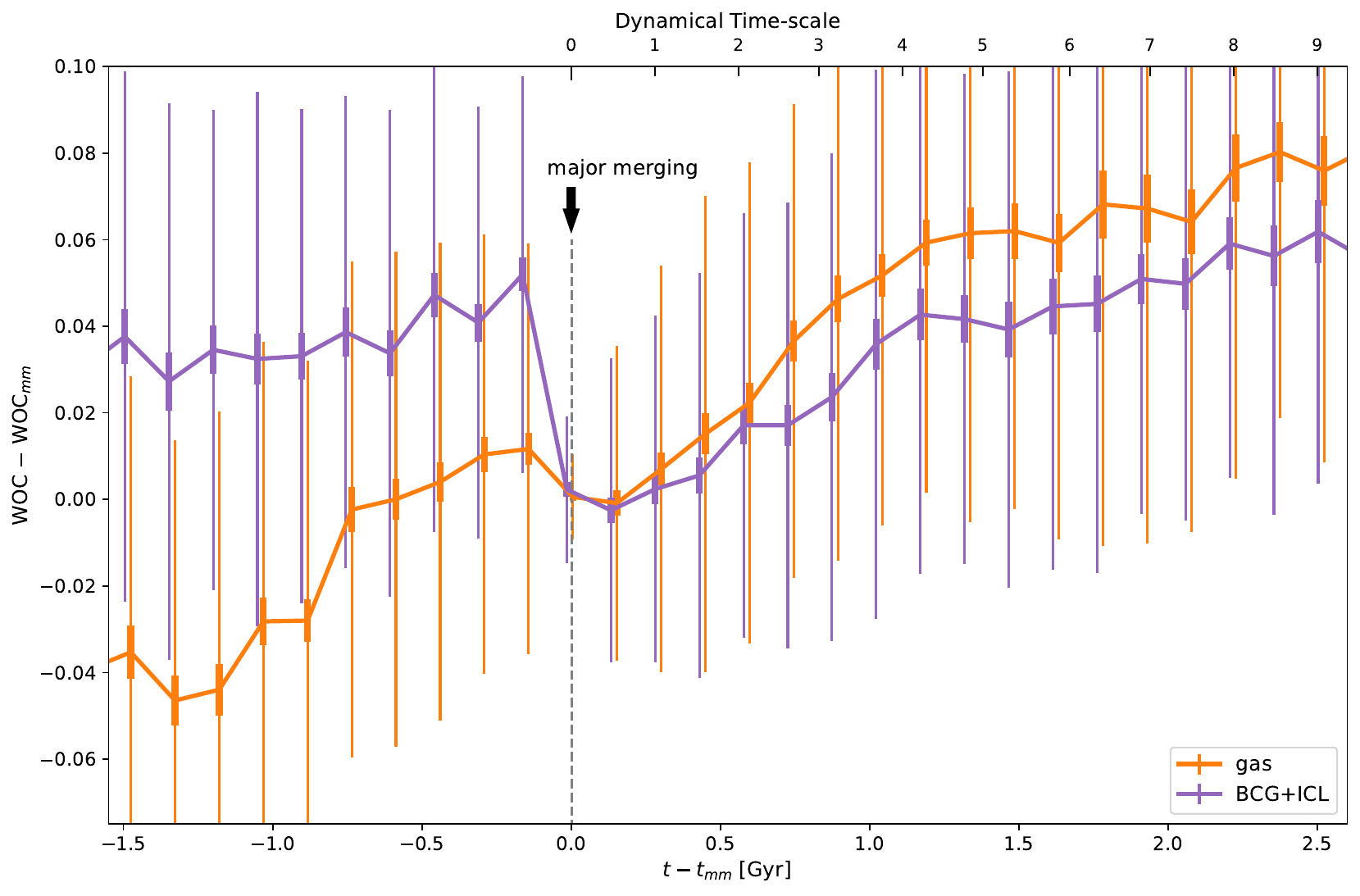}
\caption{Temporal variation of spatial distribution similarity between dark matter and gas (orange trend line) or BCG+ICL (purple trend line) before and after isolated (no other major merging within 1 Gyr) major merging (between 2 snapshots more than 33\% increase of system mass). The trend lines represent each bin's median value of $\Delta$WOC (temporal deviation of WOC value from the WOC at the major merging moment ($\mathrm{WOC}_\mathrm{mm}$)). 
Thin vertical lines represent the standard deviation, while thick vertical lines denote the standard error, and the points are shifted horizontally for clarity. The mean dynamical time scale of the cluster inner region ($< 0.3$ $r_\mathrm{vir}$) is indicated on the upper side of the $x$-axis.}
\label{fig:evo}
\end{figure}

\begin{figure}
\centering
\includegraphics[width=0.9\textwidth,trim={2cm 0 2cm 0},clip]{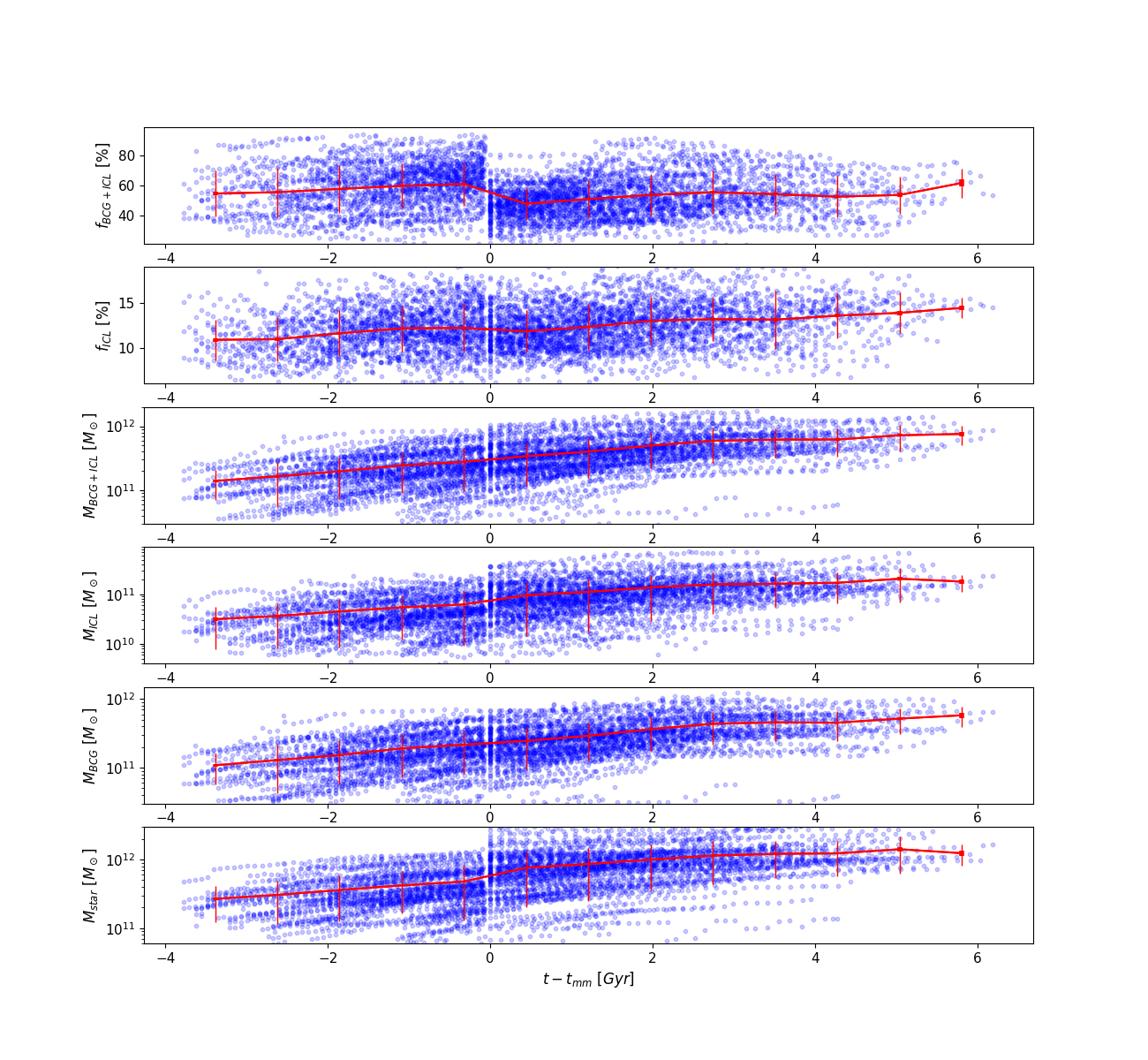}
\caption{BCG+ICL fraction, ICL fraction, BCG+ICL mass, ICL mass, BCG mass, and total stellar mass evolution before and after isolated (no other major merging within 1 Gyr) major merging (between 2 snapshots more than 33\% increase of system mass). Thin vertical lines are standard deviations,
and thick vertical lines are standard errors.}
\label{fig:fracevo}
\end{figure}

{\bf Variation of WOC around major merging}:
We hypothesize that WOC values reflect the degree of relaxation of a system by capturing the alignment between dark matter and components like BCG+ICL or gas. Our key question is whether WOC values fluctuate significantly during events that disrupt system relaxation, such as major mergers, and whether they recover post-merger.

To investigate this, we tracked WOC changes before and after major mergers, focusing on \textit{isolated major merging} events identified using the criteria described in Figure \ref{fig:massgrowth}. Specifically, we looked for mergers causing more than a 33\% mass increase within a snapshot interval and without other major mergers occurring within 1 Gyr before or after. This approach yielded 96 galaxy clusters with 117 isolated major merging events.

The WOC (DM, BCG+ICL) values were measured for each cluster at each snapshot around these 117 major merging events. The results, stacked according to their merging times, are presented in Figure \ref{fig:evo}. This figure illustrates the temporal variation of WOC values relative to the value at the merging moment. A key observation is that both BCG+ICL and gas exhibit a decrease in WOC just before the major merger and an increase afterward. In the BCG+ICL case, the WOC difference is about 0.05 at the merger time, indicating a 5\% drop in spatial distribution similarity due to the merger. The change in the gas component is less pronounced but follows a similar pattern.

This trend suggests that the approach of two massive objects and their subsequent merger temporarily disrupts the relaxation of the primary object, reflected in the lowered WOC values. The WOC values rebound as the system re-establishes equilibrium within approximately 1 to 1.5 Gyr  for BCG+ICL and 0.5 Gyr for gas, signifying a return to a more relaxed state. 
The calculated mean dynamical time-scale  (${t_\mathrm{dyn}=(2R^3/GM)^{1/2}}$) for the entire galaxy cluster is approximately 1.7 Gyr. However, the region considered for the WOC calculation extends only to 0.3 $r_\mathrm{vir}$. Assuming that most of the mass is within 0.3 $r_\mathrm{vir}$, the calculated mean dynamical time-scale for the inner region of the cluster is around 0.28 Gyr. Thus, the WOC recovery time (1 - 1.5 Gyr) corresponds to approximately 4 - 5 times the mean dynamical time-scale of the inner region of the sample clusters. 
These observations underscore the potential of the WOC as a dynamic measure of system relaxation, with its temporal changes mirroring the dynamical state of the system.

To delve deeper into the dynamics surrounding isolated major merging events, we examined changes in the BCG+ICL fraction, ICL fraction, and the masses of BCG, ICL, and BCG+ICL before and after these events (as depicted in Figure \ref{fig:fracevo}). During a major merger, defined as a moment where the total mass of the system (including dark matter) increases by more than 33\% compared to the previous snapshot, we observed a corresponding rise in the total stellar mass  (sixth panel of Figure \ref{fig:fracevo}). This is typically the case of the `sudden inclusion of a galaxy group that is heavier than 33\% of the main cluster'. Such galaxy groups would have their `brightest group galaxy (BGG)', which will not be accounted as BCG+ICL, but rather as `satellite galaxy'. In the meantime, the BCG in the main cluster is still the brightest galaxy after the inclusion of the galaxy group. Thus, the growth of BCG will follow rather gradually compared to the one snapshot interval. Therefore, the BCG mass is relatively stable, showing only modest growth over time (visible in the fifth panel of Figure \ref{fig:fracevo}). This stability in BCG mass leads to a decrease in the BCG+ICL fraction, which is calculated as the combined stellar mass of BCG and ICL divided by the total stellar mass of the cluster. 

Concurrently, the ICL fraction, representing the ICL stellar mass as a proportion of the total stellar mass, exhibited a steady increase from 10\% to 15\%. However, this increase in the ICL fraction  is not significantly impacted by major merger events. At the time of these mergers, we observe a substantial production of ICL, possibly supplemented by pre-processed intragroup light (IGL) being integrated into the ICL of the main cluster (as shown in the ICL mass graph in the fourth panel of Figure \ref{fig:fracevo}). Despite this, the ICL fraction displayed a slight downward trend, likely due to enhanced star formation within galaxies triggered by the merging processes. The typical snapshot interval of about 100 Myr is sufficient for an additional $\sim$ 10\% of stars to form \citep{park22}. The trends in the masses of BCG+ICL, ICL, and BCG, growing in a similar pattern (as seen in the third, fourth, and fifth panels of Figure \ref{fig:fracevo}), corroborate findings from a previous ICL simulation study \citep{2022ApJ...928...99C}. Moreover, the stability of the ICL fraction over approximately 10 Gyrs aligns with recent observational findings on the co-evolution of ICL mass and cluster mass \citep{2023arXiv230900671Z}.

\begin{figure}
\centering
\includegraphics[width=0.9\textwidth]{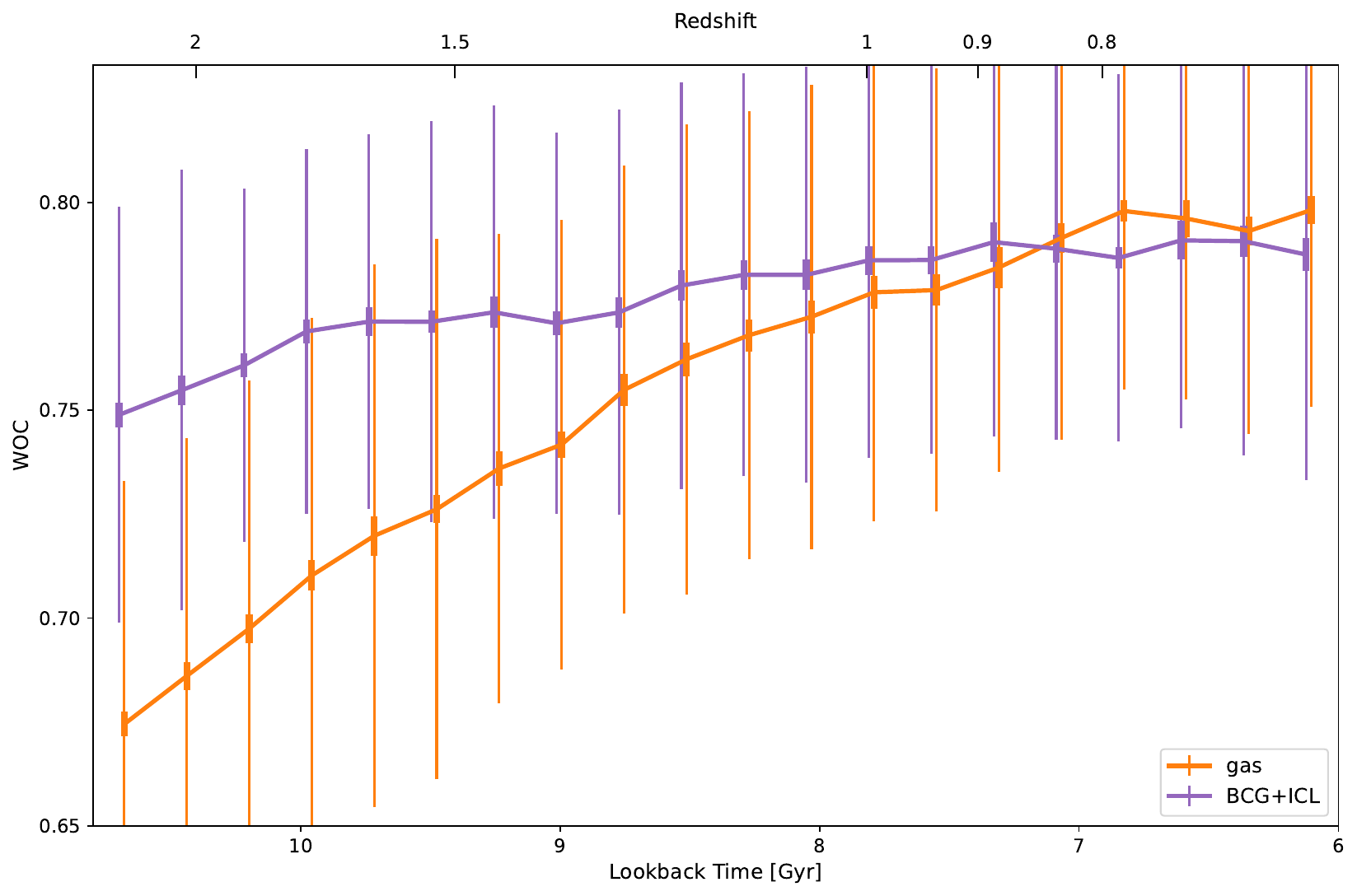}
\caption{Evolution of spatial distribution similarity between dark matter and gas (orange trend line) or BCG+ICL (purple trend line) over cosmological timescales. The same data set was used as Figure \ref{fig:evo}.
Thin vertical lines represent the standard deviation, while thick vertical lines denote the standard error, and the points are shifted horizontally for clarity. }
\label{fig:evo2}
\end{figure}

{\bf Evolution of WOC}: 
Utilizing the same dataset of WOC values from 96 galaxy clusters, as illustrated in Figure \ref{fig:evo}, we examined the broader trends in the WOC evolution, independent of major merging events. Figure \ref{fig:evo2} presents the overall temporal evolution of the spatial distribution similarity between dark matter and the BCG+ICL/gas components.

Our findings indicate notable stability in the WOC (DM, BCG+ICL), which consistently exceeds 75\% throughout the observed time interval. This suggests a persistent and strong correlation between the spatial distribution of BCG+ICL and dark matter. In contrast, the WOC (DM, gas) exhibits a gradual increase over time. Notably, at a lookback time greater than 10.4 Gyr (approximately $z \sim 2$), gas was significantly less effective in tracing dark matter than BCG+ICL. However, around a lookback time of 7 Gyr  ($z \sim 0.8$), the effectiveness of the gas component in tracing dark matter improved, eventually matching the WOC levels of BCG+ICL. This upward trend in WOC (DM, gas) corroborates the overall slope observed in Figure \ref{fig:evo} and suggests that the ability of the gas component to trace dark matter has evolved over time.

These observations imply that at higher redshifts ($z > 1$), BCG+ICL is a more reliable tracer of dark matter than the gas component. The collisionless nature of the BCG+ICL component, likely formed within the host dark matter halo, enables it to trace the dark matter distribution from an early stage. Conversely, at around $z \sim 2$ ($\sim$ 10.4 Gyr lookback time),  systems are not yet fully developed at the cluster scale, and the presence of hot gas is minimal. 
The cold gas accreted through filaments undergoes shock heating and heating from AGN jets.
Cold gas remains isothermal, undergoing shocks, while hot gas is adiabatic, behaving more closely to dark matter \citep{2003MNRAS.345..349B, 2009Natur.457..451D}.
Over time, as the gas within the cluster predominantly turned hot, with only a minor fraction remaining cold within galaxies, it gradually approached hydrodynamic equilibrium. This process allowed the gas to eventually become a comparably effective dark matter tracer at later times ($z < 0.9$).

\subsection{Radial profiles}
In addition to two-dimensional comparisons, we assessed the spatial distribution similarities of different galaxy cluster components using one-dimensional radial profiles. This approach complements previous studies like \citet{2021MNRAS.501.1300S}, which also utilized one-dimensional radial profile comparisons, particularly for ICL and dark matter.

Our radial profile analysis procedure is as follows;
Firstly, the radial profiles were computed by taking azimuthal averages of the density field for each cluster component. Then, the radial profiles of the components of 174 galaxy clusters were mean stacked. The three different projection maps were used as if they were separate clusters; thus, 174 $\times$ 3 radial profiles were mean stacked. We chose mean stacking over median stacking because many simulated cluster images predominantly feature zero pixel values at outer radii, which could skew the radial profiles. The mean stacking method is more accurate in representing the actual mass or light distribution.

\begin{figure}
\centering
\includegraphics[height=6cm,trim={0.1cm 0 0.1cm 0.1cm},clip]{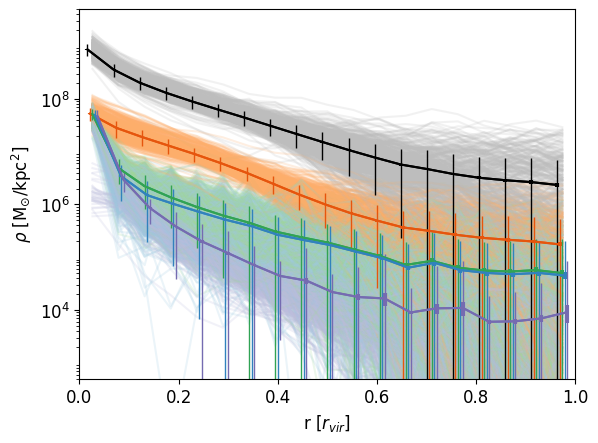}
\includegraphics[height=6cm,trim={0.1cm 0 0.1cm 0.1cm},clip]{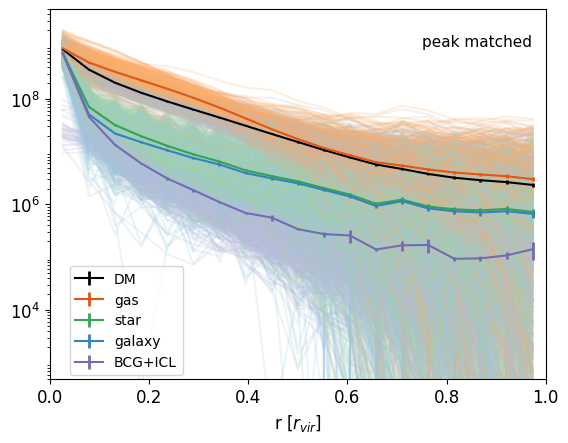}
\caption{The radial profiles of various components of 174 galaxy clusters. The three orthogonal projection angles are regarded as individual galaxy clusters, which makes the number of samples 522. Bold lines denote the mean stacked, whereas the dimmer thin lines are individual radial profiles of the corresponding color components. The virial radius of the galaxy cluster normalizes each radial profile. (Left) Amplitudes represent the density of each radial profile.  Thin vertical lines represent the standard deviation, while thick vertical lines denote the standard error, and the points are shifted horizontally for clarity. (Right) The peak amplitudes of each radial profile are scaled to match the dark matter one. The vertical lines denote the standard error. For the scale of standard deviation, refer to the left panel.}
\label{fig:profile174_mean_peak2}
\end{figure}
\begin{figure}
\centering
\includegraphics[width=0.9\textwidth,trim={0.2cm 0.2cm 0 0.1cm},clip]{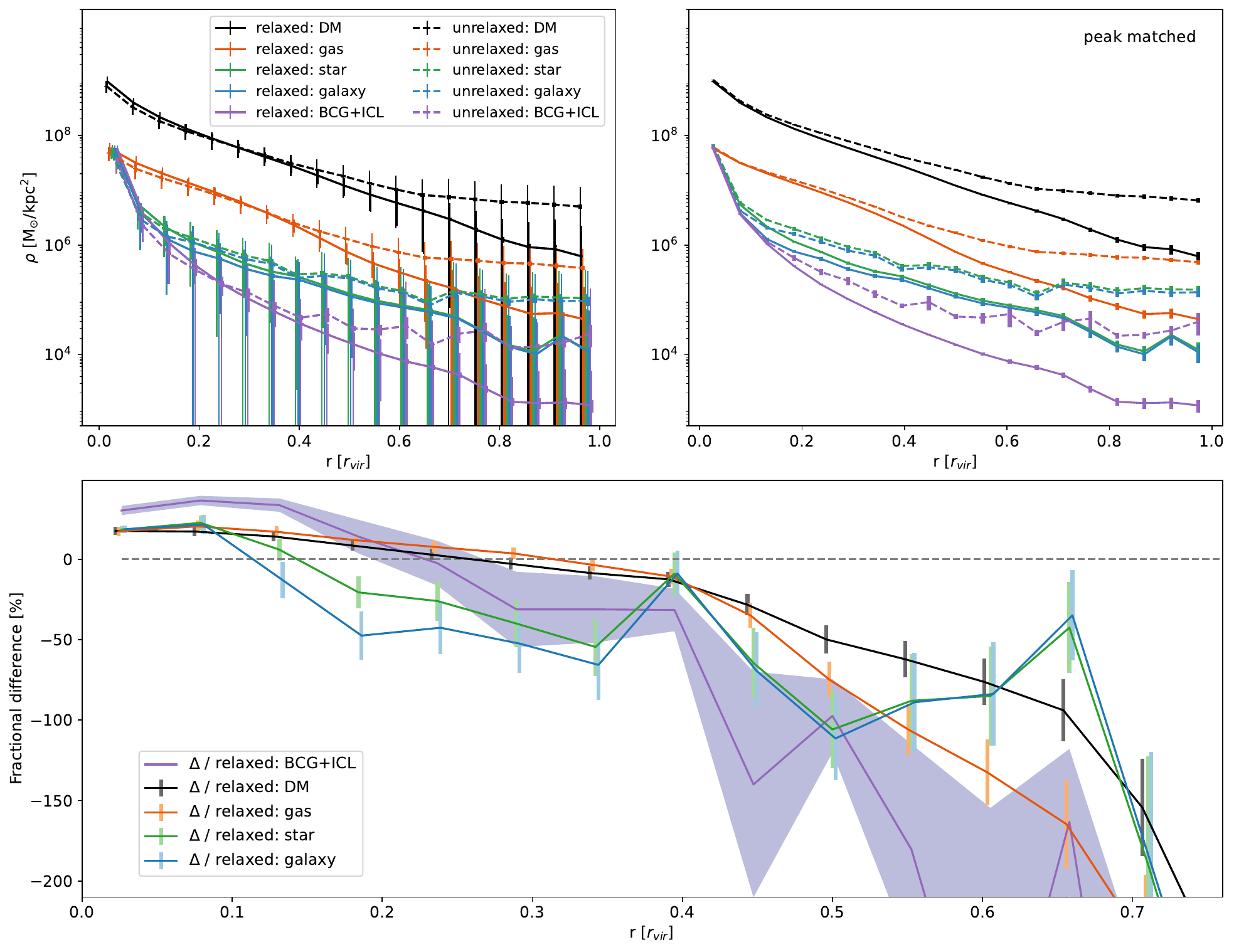}
\caption{(Upper left) The stacked radial profiles of various components of the 55 most relaxed/ the 54 most unrelaxed galaxy clusters. Solid lines denote the components of relaxed clusters, whereas dashed lines devote the components of unrelaxed clusters. Amplitudes represent the density of each radial profile. Thin vertical lines represent the standard deviation, while thick vertical lines denote the standard error, and the points are shifted horizontally for clarity.  (Upper right) The peak amplitudes of each unrelaxed radial profile are scaled to match the relaxed one. Notations are identical to the upper left subplot. The vertical lines denote the standard error. For the scale of standard deviation, refer to the left panel. (Bottom) The residual of relaxed and unrelaxed profiles normalized by the relaxed one. The error propagations of the standard errors are shown as error bars  and shifted horizontally for clarity.}
\label{fig:profile_mean_mean}
\end{figure}

The analysis showed that the radial profile of the gas distribution closely resembles that of dark matter. In contrast, the profiles of stars and galaxies appeared similar to each other but diverged from the BCG+ICL and dark matter profiles (as illustrated in the right panel of Figure \ref{fig:profile174_mean_peak2}). These observations align with \citet{2021MNRAS.501.1300S}, highlighting the concentrated nature of the diffuse stellar component. Additionally, our findings support the conclusions of \citet{2022ApJS..261...28Y}, demonstrating that while WOC is insensitive to concentration rates, it effectively captures the two-dimensional spatial distribution similarities.

Interestingly, the one-dimensional radial profiles of stars and galaxies are quite similar, differing notably from those of BCG+ICL and gas, and only the gas profile seems comparable to that of dark matter. However, when examining the two-dimensional spatial distributions using the WOC method, we find that both BCG+ICL and gas trace dark matter effectively, with WOC values of 0.792 and 0.813, respectively. In contrast, stars are less effective in tracing dark matter (WOC = 0.661), and galaxies are even less so (WOC = 0.557). 
This finding highlights the criticality of employing two-dimensional spatial distribution analyses, in conjunction with one-dimensional radial profile studies, for a more comprehensive understanding of the spatial correlations with dark matter.

Next, we divided the sample galaxy clusters using the half-mass epoch. The most early-formed 55 clusters are categorized as relaxed, and the most late-formed 54 clusters are classified as unrelaxed systems.
We stacked the radial profiles of 55 relaxed and 54 unrelaxed galaxy clusters for their various components: dark matter, stars, galaxies, BCG+ICL, and gas. We checked how the radial profile of the components of the relaxed and  unrelaxed (merging) cluster differs (see Figure \ref{fig:profile_mean_mean}).
Again, the three different projection maps were used as if they were other clusters; thus, each radial profile results from the mean stacking of 55 (54) $\times$ 3 radial profiles.
Every time, the center is fixed as the peak of the dark matter, and the virial radius of the cluster normalizes the radial bins.

In the upper panel of Figure \ref{fig:profile_mean_mean}, dark matter, gas, stars, and BCG+ICL of unrelaxed clusters (dashed lines) seem more broadly spread than relaxed clusters (solid lines). In the inner region (0 $\sim$ 0.2 $r_{\text{vir}}$) in galaxy clusters, the relaxed samples show higher density, whereas in the outer part (0.6 $\sim$ 1 $r_{\text{vir}}$), the unrelaxed samples show higher density (see the lower panel of Figure \ref{fig:profile_mean_mean}). These results may be related to the fact that the components in the unrelaxed samples are scattered out during the active merging events. In contrast, the relaxed sample gathered their components inward and built more concentrated profiles. The lower panel shows the  difference of relaxed and unrelaxed cluster profiles normalized by relaxed profiles. We could see clearly the exceeded density of relaxed clusters in the inner radius (above zero) and unrelaxed clusters in the outer radius (below zero). Interestingly, the BCG+ICL (purple line) shows the most significant deviation between relaxed and unrelaxed cases, which tells us again that BGC+ICL is a sensitive proxy for dynamical history. In the inner region (0 $\sim$ 0.2 $r_{\text{vir}}$), the radial density of BCG+ICL differs up to $\sim$ 40\% for relaxed and unrelaxed cases.  

{\bf Recovering dark matter radial profile from BCG+ICL profile}:
In Section \ref{subsec:result1}, we primarily focused on two-dimensional spatial comparisons, setting aside the direct relationship between the field strengths. However, if our goal is to find a component that can represent the radial profile of dark matter, this approach may need refinement. Prior studies comparing the profiles of dark matter and ICL (diffuse light component) reported varying slopes, typically showing the ICL profile to be more radially concentrated and thus not perfectly mirroring the dark matter profile. Despite this, the ICL profile slope has been found to correlate strongly with the total mass of the cluster \citep{2014MNRAS.444..237P, 2018MNRAS.475..648P, 2018MNRAS.474..917M, 2021MNRAS.501.1300S}.
As shown in Figure \ref{fig:profile174_mean_peak2}, the dark matter profile decreases 
gradually, contrasting with the steeper BCG+ICL profile.
This led us to explore methods for deriving the dark matter profile from the BCG+ICL profile.

\begin{figure}
\centering
\includegraphics[width=0.5\textwidth]{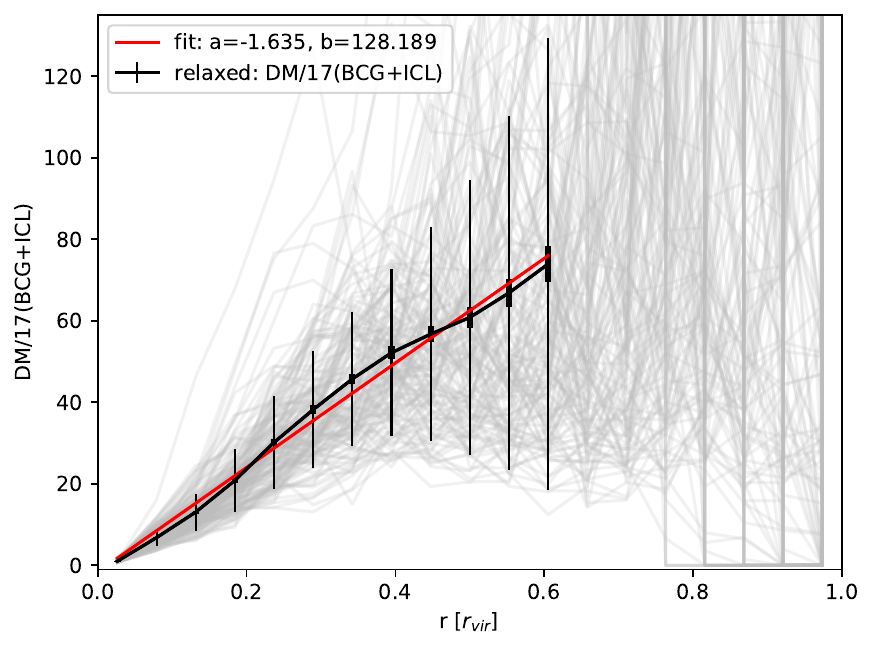}
\caption{Ratio between dark matter and scaled BCG+ICL profiles of 55 relaxed galaxy clusters (gray lines), mean value and standard error (black solid line), and linear ($a+br$) 
 fitting function (red line). The three orthogonal projection angles are regarded
as individual galaxy clusters, which makes the number of samples 165. Thin vertical lines represent the standard deviation, while thick vertical lines denote the standard error.}
\label{fig:dm_radial}
\end{figure}

In pursuit of the most robust solution, we considered the 55 relaxed galaxy clusters exclusively, each observed from three distinct orthogonal projection angles, resulting in a total sample size of 165. Much like the right panel of Figure \ref{fig:profile174_mean_peak2}, we scaled the BCG+ICL profile by a factor of approximately 17 to align with its peak value. Subsequently, we divided the dark matter profile by the scaled BCG+ICL profile for all 165 galaxy cluster samples, represented by the gray lines in Figure \ref{fig:dm_radial}. Notably, the mean profile of this dark matter - BCG+ICL ratio (black solid line) appears to exhibit an approximate linear trend with respect to the radial distance from the cluster center. 
The ratio exhibits significant noise beyond 0.6 virial radii due to the limited presence of ICL in that region. Our approach involved fitting the ratio within the range of 0 $\sim$ 0.6 $r_{\text{vir}}$, where we employed a linear fitting function ($a+br$), represented as a continuous red line. Such a clear trend was not observed in analyses involving gas or stars (BCG+ICL+satellite galaxies).

Our findings suggest the potential for reconstructing the dark matter profile from the BCG+ICL profile by applying specific scaling constants and considering the radial distance from the center of the cluster. Conversely, with an established dark matter distribution from weak-lensing or N-body simulations, we can estimate the BCG+ICL radial profile or its extent. It is intriguing to note that the quantity BCG+ICL exhibits an inverse radius scaling relationship ($\sim 1/r$) relative to dark matter. This phenomenon could potentially be attributed to ``mass segregation'', where the more massive galaxies tend to concentrate in the inner region of the cluster, while lighter galaxies are more prevalent in the outer region. This segregation could lead to a higher presence of ICL in the inner radius compared to the outer radius. 

A caveat pertinent to the scaling relationship between the profiles of dark matter and BCG+ICL might be warranted, considering that the BCG predominantly influences the BCG+ICL profile. The precise nature of this relationship could potentially vary based on the adiabatic cooling and feedback recipes employed in the simulation. 

\section{Discussions} \label{sec:discusion}
Our study has focused on identifying the galaxy cluster components that most effectively trace dark matter and examined how their tracing capabilities vary with the dynamical state of the cluster. We also  investigated methods to improve dark matter tracing, aiming to achieve higher WOC results.

\begin{figure}
\centering
\includegraphics[width=0.9\textwidth]{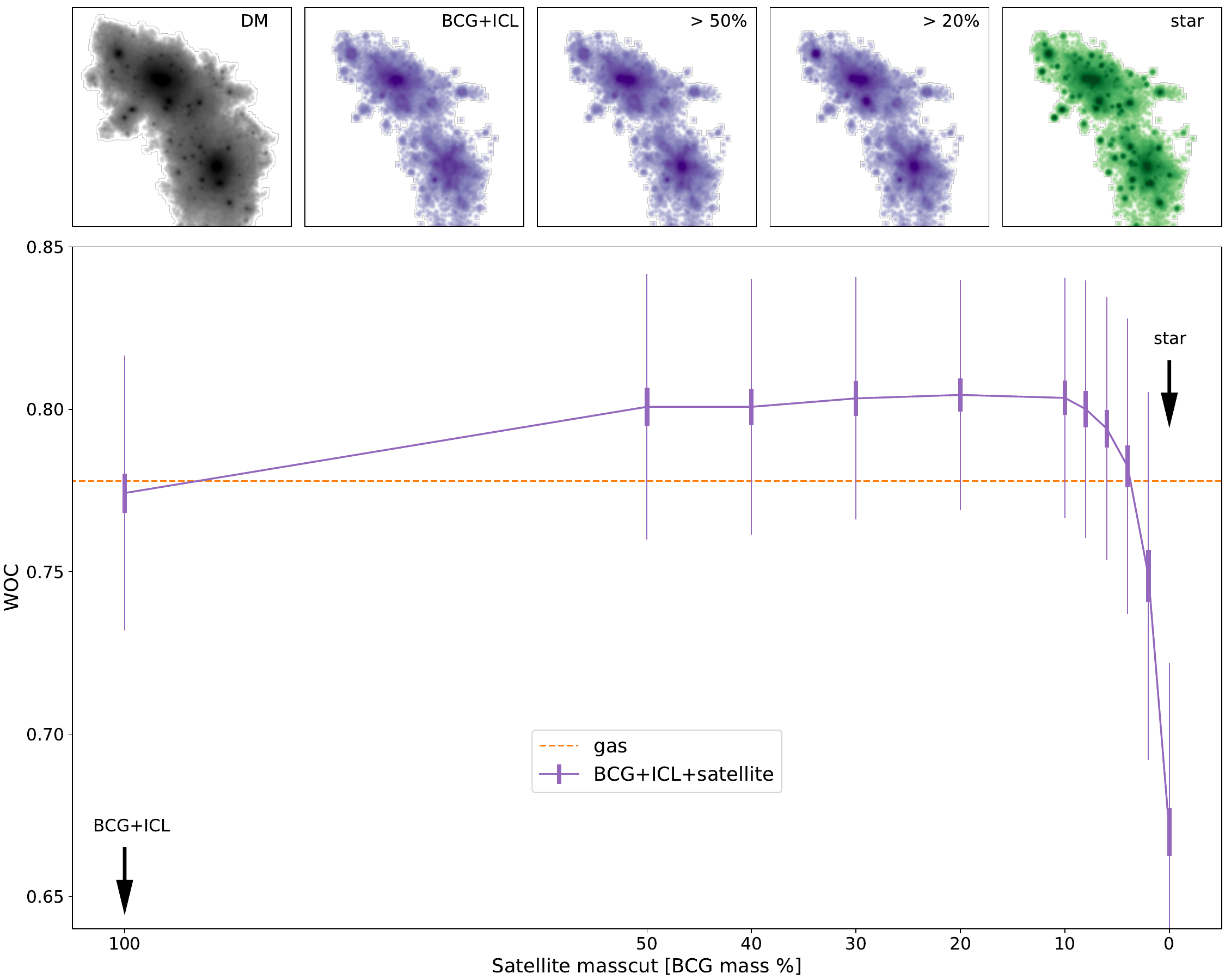}
\caption{(Top) 2D projected images of an example galaxy cluster (Unrelaxed in Figure \ref{fig:example2}). The images cover one virial radius of the cluster. Dark matter, BCG+ICL, BCG+ICL+satellite galaxies heavier than 50\%, 20\% of BCG mass, and all stellar particles are shown respectively from left to right. (Bottom) WOC changes upon the inclusion of satellite galaxies with different mass cuts. It is calculated using 51 galaxy clusters at the final snapshot ($z=0.625$), which contains satellite galaxy heavier than 50\% of BCG stellar mass. Thin vertical lines represent the standard deviation, while thick vertical lines denote the standard error. WOC peaks at the inclusion of satellite galaxies above 20\% of the BCG mass.}
\label{fig:masscut}
\end{figure}

{\bf Recipe for tracing dark matter in unrelaxed system}: 
An ideal dark matter tracer should be collisionless and effective even in disturbed systems. Observations of the Bullet cluster, a classic example of a disturbed system, show a spatial mismatch between dark matter and hot gas, while the collisionless stars align more closely with the dark matter distribution.
In Figure \ref{fig:evo}, a noticeable drop in WOC (DM, BCG+ICL) is observed around major merging events. This drop can be attributed to the merging of new galaxy groups, each with its own BGG. In the newly formed main cluster, these BGGs are categorized as satellite galaxies and are excluded from the BCG+ICL component, creating a discrepancy in the spatial distribution, particularly where significant amounts of dark matter are present. This phenomenon likely contributes to the approximately 5\% reduction in WOC (DM, BCG+ICL) at major merging points. Additionally, this explains the observed relationship between the relaxedness of the system and the WOC (DM, BCG+ICL) shown in Figure \ref{fig:woc_dyn}, as unrelaxed systems often contain larger satellite galaxies with substantial amounts of dark matter.

We tested whether including large mass satellite galaxies into the BCG+ICL component improves the spatial distribution similarity with dark matter. 
Among the 174 galaxy clusters at the final snapshot, we found 51 clusters containing satellite galaxies heavier than 50\% of the BCG in stellar mass. These clusters would be mostly unrelaxed, having a small magnitude gap ($\Delta M_{12}$) as well as a high substructure fraction (see Section \ref{subsec:dynamic}). We then measured the similarity between the dark matter and the BCG+ICL+satellite galaxies, which fulfill certain mass criteria.
Figure \ref{fig:masscut} shows how the WOC value changes upon inserting satellite galaxies with various mass cuts. 

The orange dashed line indicates the WOC (DM, gas) of the same 51 sample clusters in the final snapshot. The BCG+ICL alone (which corresponds to the mass criteria as 100\% of BCG mass) shows a slightly lower WOC value than that of gas. As we include satellite galaxies that are heavier than 50\% of the BCG mass, we observe that it improves the WOC value and becomes higher than the WOC (DM, gas) value. We include more satellite galaxies by decreasing the mass cut criteria 40\%, 30\%, 20\%, 10\%, 8\%, 6\%, 4\%, and 2\%. The WOC value peaks at the inclusion of satellite galaxies heavier than 20\% of BCG mass and decreases rapidly after the inclusion of 10\%. If we include small satellite galaxies such as 2\% of BCG mass, the similarity with dark matter is broken, and the WOC becomes lower than the WOC (DM, gas). Finally, lowering the mass cut to 0\% means including every satellite, equivalent to all stellar particles (BCG+ICL+satellite galaxies), resulting in a much worse WOC value.

As we recognized in Section \ref{subsec:result1}, taking account of all satellite galaxies is not ideal for tracing the dark matter in the system. These smaller satellite galaxies may have migrated from outside through filaments and would not have enough time to blend into the global gravitational potential of the cluster. Moreover, we observed some small satellite galaxies without dark matter in the simulation.  In contrast, the BCG and ICL have developed within the dark matter halo of the cluster, aligning their distributions more closely with dark matter. Thus, in addition to the BCG+ICL, we suggest including massive satellite galaxies, which are heavier than 20\% of BCG mass, to trace dark matter, particularly in disturbed galaxy clusters.

{\bf Tracing dark matter combining BCG+ICL and gas}:
As highlighted in Section \ref{subsec:result1}, at the final snapshot (z=0.625), both BCG+ICL and gas components were found to be effective tracers of dark matter, each exhibiting approximately 80\% spatial distribution similarity. When observational data for both components are available, combining them could potentially enhance our ability to trace dark matter with greater fidelity. Furthermore, as observed in Figure \ref{fig:example2}, the dark matter distribution appears to be intermediate between the BCG+ICL and gas distributions. While the gas distribution is considerably smoother, the BCG+ICL displays more detailed substructures than observed in the dark matter.

We integrated the BCG+ICL and gas density maps to evaluate this combined approach, normalizing each based on their respective total masses to ensure equal representation. The findings of the resultant WOC (DM, BCG+ICL+gas) are presented in the last row of Table \ref{tab:table2}. This integration shows an improvement in tracing the spatial distribution of dark matter: while the individual similarity percentages are 79.2\% for BCG+ICL and 81.3\% for gas, the combined approach yields an enhanced similarity of 86.2\%. This enhancement is consistent across various dynamical states of the clusters. In the case of relaxed clusters, the combined BCG+ICL and gas distribution traces dark matter with an approximate accuracy of 88.7\%.

These results indicate that a synergistic approach, utilizing both BCG+ICL and gas components, provides a more comprehensive method for tracing the spatial distribution of dark matter within galaxy clusters. This methodology benefits from the distinct characteristics of each component, leading to a more accurate representation of the dark matter distribution compared to using either component alone.

\section{Conclusions} \label{sec:conclusion}
We have conducted an extensive comparative study of the spatial distributions of various components within simulated galaxy clusters. We used the Horizon Run 5 simulation, focusing on 174 high-mass clusters ($> 5 \times 10^{13} M_{\odot}$) at $z=0.625$. We used the Weighted Overlap Coefficient (WOC) method, a novel approach to quantifying the similarity of two-dimensional spatial distributions.

We measured the spatial distribution similarity between dark matter and various components such as all stellar particles, all galaxies, BCG+ICL, and gas. We also investigated if the similarity has any relation with the dynamical state of the cluster. Our findings suggest that the spatial distribution of the BCG+ICL and the gas coincide with the dark matter distribution, particularly for more relaxed galaxy clusters (see Table \ref{tab:table2}, Figure \ref{fig:woc_dyn}). 
Our study shows that the ICL, when combined with the BCG, can serve as a luminous tracer for dark matter, implying that the spatial distribution similarity against the dark matter may be used as a probe of the dynamical state of the cluster. However, for the application to individual galaxy clusters, we should be cautious, due to the large uncertainty on the measurement.

Focusing on clusters experiencing major merging, we found that the spatial similarity between dark matter and BCG+ICL dropped just before major merging and recovered 1 - 1.5 Gyr after (see Figure \ref{fig:evo}). This suggests again that the WOC (DM, BCG+ICL) could act as a probe of the dynamical state of the system. Moreover, considering the overall evolution of the WOC over cosmological timescales, particularly for galaxy clusters at redshift greater than 1, the BCG+ICL seems to be a better tracer for dark matter than gas (see Figure \ref{fig:evo2}). From an observational point of view, particularly at higher redshift, observing the BCG+ICL using deep imaging  could be more advantageous than weak-lensing analyses or X-ray observations, which suffer from various systematic. Considering the observational study of ICL at z $\sim$ 0.5 \citep{2021MNRAS.508.2634Y}, we may need at least 29 mag/arcsec$^2$ deep detection limit to explore the ICL out to 0.3 $r_{\text{vir}}$.

Additionally, the one-dimensional radial profiles demonstrated that the BCG+ICL is a sensitive probe of the dynamical state of galaxy clusters ( see the lower panel of Figure \ref{fig:profile_mean_mean}).
We further suggest a method to retrieve the  approximate dark matter profile, in relaxed clusters, by scaling the BCG+ICL profile (see Figure \ref{fig:dm_radial}).

Furthermore, we suggest a recipe for tracing dark matter in unrelaxed systems by including satellite galaxies with stellar masses greater than 20\% of the BCG mass (see Figure \ref{fig:masscut}). Additionally, if both BCG+ICL and gas data are available, we could combine them to trace the spatial distribution of the dark matter with higher fidelity (see Table \ref{tab:table2}). These results may guide future observational studies and inform the selection of tracers to study galaxy cluster evolution. 

We may speculate as to whether our results will hold at z=0. In Figure \ref{fig:fracevo} the amount of ICL increases through a major merger, while in Figure \ref{fig:evo}, the WOC(DM, BCG+ICL) is affected through the major merger but recovers in $\sim$ 1Gyr. We expect the BCG+ICL, as a collisionless component, will continue to trace DM as well as gas, which has already reached hydrodynamic equilibrium. Even after the burst of ICL production at $z<$0.6 \citep{2022NatAs...6..308M}, probably related to merging events, the temporarily reduced WOC value will recover in 1 $\sim$ 1.5 Gyr. However, this should be rigorously tested in the near future with simulations at lower redshift.

Future work should examine more diverse samples of galaxy clusters spanning various masses and redshifts and compare them with observational studies. Recent deep imaging surveys are yielding statistical insights into ICL properties, including color and color gradient \citep{2023arXiv230900671Z,2023MNRAS.521..478G}, extending observations up to redshift $z\sim 2$ \citep{2023MNRAS.523...91W}. 
It would also be interesting to examine the origin of ICL depending on the mass accretion history and angular momentum of the galaxy cluster, as well as the environment surrounding the cluster as a node of the large-scale structure filaments. Further analysis of the \hr\ simulation could give us hint on the relation between the initial condition of the local density field and structure formation history at late times.
Improvements in our understanding of the ICL and its relation to dark matter will be aided by forthcoming observational facilities and instruments. As observations continue to improve, testing and validating these results in real-world observations of galaxy clusters will be crucial.

\begin{acknowledgments}
We would like to thank the anonymous referee whose careful and insightful comments allowed us to significantly improve the presentation of this work. 
J.Y. was supported by a KIAS Individual Grant (QP089901) via the Quantum Universe Center at Korea Institute for Advanced Study.
C.G.S is supported via the Basic Science Research Program from the National Research Foundation of South Korea (NRF) funded by the Ministry of Education (2018R1A6A1A06024977 and  2020\-R1\-I1\-A1\-A01073494). A.S., C.P. and J.K. are supported by KIAS Individual Grants (PG080901, PG016903, KG039603) at Korea Institute for Advanced Study. J.L. is supported by the National Research Foundation of Korea (NRF-2021R1C1C2011626). J.K. and J.L. are supported by the National Research Foundation of Korea (NRF) grant funded by the Korea government (MSIT, 2022M3K3A1093827). M. J. J. acknowledges support for the current research from the NRF of Korea under the programs 2022R1A2C1003130 and RS-2023-00219959. O.S. acknowledges support from an ERC Consolidator Grant (Grant Agreement ID 101003096) and STFC Consolidated Grant (ST/V000721/1). Y.K. is supported by Korea Institute of Science and Technology Information (KISTI) under the institutional R\&D project (K-24-L02-C04). This work benefited from the outstanding support provided by the KISTI National Supercomputing Center and its Nurion Supercomputer through the Grand Challenge Program (KSC-2018-CHA-0003, KSC-2019-CHA-0002). 
Large data transfers were supported by KREONET, which is managed and operated by KISTI. This work is supported by the Center for Advanced Computation at Korea Institute for Advanced Study. 
\end{acknowledgments}

\vspace{5mm}
\facilities{KIAS Linux Cluster Systems (Baekdu)}

\software{{\bf Astropy \citep{2013A&A...558A..33A,2018AJ....156..123A,2022ApJ...935..167A}}
          }

\bibliography{HR5}{}
\bibliographystyle{aasjournal}

\end{document}